\documentclass[12pt]{article}
\usepackage{jheppub_modified_green}

\usepackage{epsfig}
\usepackage{amsfonts}
\usepackage{amssymb,esint}
\usepackage{enumitem}
\usepackage{graphicx}
\usepackage{amsthm}
\usepackage{subfig}
\usepackage{bm}
\usepackage{hyperref}
\usepackage{mathrsfs}
\usepackage{bbm}
\usepackage{cancel}
\usepackage{comment}
\usepackage{xcolor}
\usepackage{relsize}
\usepackage{float, slashed, graphicx, amssymb, amsmath}
\usepackage{shuffle}
\usepackage{pgfplots}
\usepackage{tikz-feynman}
\tikzfeynmanset{compat=1.1.0}
\tikzfeynmanset{graviton/.style={circle, draw=green!60, fill=green!5, very thick, minimum size=7mm}}
\tikzfeynmanset{codot/.style={/tikz/shape=circle,/tikz/fill=white,/tikz/minimum size=0.1cm,/tikz/inner sep=1.8pt}}
\tikzfeynmanset{myblob/.style={/tikz/shape=rectangle,/tikz/fill=red,/tikz/minimum size=0.2cm,/tikz/inner sep=1.8pt} }
\tikzfeynmanset{ghc/.style={/tikz/shape=circle,/tikz/fill=white,/tikz/minimum size=0.05cm,} }
\tikzfeynmanset{HV/.style={/tikz/shape=circle,/tikz/fill={rgb:black,1;white,2},/tikz/minimum size=0.1cm,} }
\tikzfeynmanset{GR/.style={/tikz/shape=ellipse,/tikz/fill={rgb:black,1;white,2},/tikz/minimum size=0.3cm,} }
\tikzfeynmanset{GR2/.style={/tikz/shape=ellipse,/tikz/fill={rgb:black,1;white,2},/tikz/minimum width = 1.2cm, 
    /tikz/minimum height = 3.4cm} }
\tikzset{box/.pic={\filldraw[fill=black]  (0,0) circle (2.5pt); \filldraw [fill=black] (0.5,0) circle (2.5pt); \draw [line width=5pt] (0,0) -- (0.5,0);}}
\usetikzlibrary{arrows.meta} 
\usetikzlibrary{calc}
\usetikzlibrary{decorations.pathmorphing}
\usetikzlibrary{decorations.pathreplacing}
\usetikzlibrary{decorations.markings}
\tikzset{
   vector2/.style={decorate, decoration={snake, amplitude=1pt, segment length=6pt}, draw,double},
   vector/.style={decorate, decoration={snake, amplitude=1pt, segment length=6pt}, draw},
	provector/.style={decorate, decoration={snake,amplitude=2.5pt}, draw},
	antivector/.style={decorate, decoration={snake,amplitude=-2.5pt}, draw},
    fermion/.style={draw=black, postaction={decorate},
        decoration={markings,mark=at position .55 with {\arrow[draw=black]{>}}}},
    fermionbar/.style={draw=black, postaction={decorate},
        decoration={markings,mark=at position .55 with {\arrow[draw=black]{<}}}},
    fermionnoarrow/.style={draw=black},
    gluon/.style={decorate, draw=black,
        decoration={coil,amplitude=4pt, segment length=5pt}},
    scalar/.style={dashed,draw=black, postaction={decorate},
        decoration={markings,mark=at position .55 with {\arrow[draw=black]{>}}}},
    scalarbar/.style={dashed,draw=black, postaction={decorate},
        decoration={markings,mark=at position .55 with {\arrow[draw=black]{<}}}},
    scalarnoarrow/.style={dashed,draw=black},
    electron/.style={draw=black, postaction={decorate},
        decoration={markings,mark=at position .55 with {\arrow[draw=black]{>}}}},
	bigvector/.style={decorate, decoration={snake,amplitude=4pt}, draw},
}
\tikzset{cross/.style={cross out, draw, 
         minimum size=2*(#1-\pgflinewidth), 
         inner sep=0pt, outer sep=0pt}}

\tikzstyle{block} = [draw, rectangle, 
    minimum height=3em, minimum width=6em]

\usepackage[T1]{fontenc} 

\newcommand{\tr}{\text{tr}}

\def\sc#1{\overline{#1}}
\makeatletter
\newcommand \UPlus {\mathop {\operator@font \uplus }\limits }
\makeatother
\makeatletter
\newcommand \Bigcup {\mathop {\operator@font \bigcup }\limits }
\makeatother
  \def\LabelNote#1{}
 \def\Label#1{\label{#1}%
  \smash{\hbox to\phipt{\raise1ex\hbox{\tiny[#1]}\hss}}}
  
  \def\mdot{{\cdot}}

\newcommand{\arccosh}{{\rm arccosh}}

\newcommand{\terms}[1]{\mathcal{O}\left(#1\right)}

\newcommand{\cM}{\mathcal{M}}

\newcommand{\eps}{\epsilon}
\newcommand{\vareps}{\varepsilon}

\def\lm{\frac{d^D\ell_1}{\pi^{D/2}}\frac{d^D\ell_3}{\pi^{D/2}}}
\def\qm{\frac{d^{D}q}{(2\pi)^{D-2}}}

\def\nn{\nonumber}

\def\spa#1.#2{\left\langle#1\,#2\right\rangle}
\def\spb#1.#2{\left[#1\,#2\right]}

\def\be{\begin{equation}}
\def\ee{\end{equation}}
\def\bea{\begin{eqnarray}}
\def\eea{\end{eqnarray}}  
\allowdisplaybreaks

\newcommand{\npre}{\mathcal{N}}

\usepackage{amssymb,amsmath}
\usepackage{mathtools} 
\usepackage{cancel} 
\usepackage{graphicx} 

\usepackage{hyperref}
\hypersetup{
	colorlinks=true,
	linktoc=page,
	citecolor=americanrose,
	linkcolor=cadmiumgreen,
	urlcolor=blue} 
\definecolor{americanrose}{rgb}{1.0, 0.01, 0.24}
\definecolor{cadmiumgreen}{rgb}{0.0, 0.42, 0.24}

\usepackage[colorinlistoftodos]{todonotes}

\def\nn{\nonumber}

\begin{document}

\title{Resumming Spinning Black Hole Dynamics at Third Post-Minkowskian Order}
\author{N. E. J. Bjerrum-Bohr$\mbox{}^{1,2}$,}
\author{Gang Chen$\mbox{}^{1,2}$,}
\author{Konstantinos Papadimos$\mbox{}^{1,2}$,}
\author{Yuexiang Zhang$\mbox{}^{2}$}
\affiliation{$^1$Center of Gravity, Niels Bohr Institute, University of Copenhagen,\\
Blegdamsvej 17, DK-2100 Copenhagen, Denmark}
\affiliation{$^2$Niels Bohr International Academy, Niels Bohr Institute, University of Copenhagen,\\
Blegdamsvej 17, DK-2100 Copenhagen, Denmark}

\emailAdd{bjbohr@nbi.dk}
\emailAdd{gang.chen@nbi.ku.dk}
\emailAdd{mqv521@alumni.ku.dk}
\emailAdd{yuexiangzhang2025@outlook.com}

\abstract{We investigate the relativistic scattering of spinning black holes using modern amplitude methods within a heavy-mass effective field theory formalism at third post-Minkowskian order. Using a systematic self-force expansion up to first order in the mass ratio, the gravitational amplitude and the associated eikonal-like phase are computed for a spin-aligned binary system comprising a heavy and a light black hole up to fifth order in the total spin and up to quadratic order in the spin of the light black hole. We also consider the resummation of the heavy black hole's spin in both the probe limit and the radiation-reaction sector, and verify that the resulting phase displays the characteristic ring singularity features associated with the Kerr metric.}

\vspace{-2.6cm}

\maketitle

\flushbottom
 \tableofcontents
%
\section{Introduction}
The relativistic two-body problem in general relativity remains a critical theoretical challenge, one that has become increasingly urgent with the rise of gravitational-wave astronomy. The detection of gravitational waves from coalescing black holes and neutron stars by the LIGO-Virgo collaboration~\cite{LIGOScientific:2016aoc} has started a new era of precision astrophysics, which necessitates a 
precise theoretical understanding of compact binary dynamics. While traditional approaches have achieved substantial success, they encounter significant computational difficulties at high perturbative orders, emphasized when spin effects---essential for accurate waveform modeling---are included.
In recent years, a transformative approach has emerged by reframing classical gravitational dynamics within the framework of modern scattering-amplitude techniques, inspired by particle physics approaches to gravity ~\cite{Iwasaki:1971vb,Donoghue:1994dn,Bjerrum-Bohr:2002gqz,Holstein:2004dn,Holstein:2008sx,Neill:2013wsa,Bjerrum-Bohr:2013bxa,Damour:2017zjx}. This relativistically valid post-Minkowskian (PM) framework organizes calculations of observables solely in powers of Newton's constant \(G_N\) rather than expansions that also involve velocities. This shift in paradigm has enabled systematic, efficient, and gauge-invariant computations of both conservative dynamics \cite{Bjerrum-Bohr:2018xdl,Cheung:2018wkq,Cristofoli:2019neg,Bern:2019nnu,Bern:2019crd,Parra-Martinez:2020dzs,DiVecchia:2020ymx,DiVecchia:2021ndb,DiVecchia:2021bdo,Herrmann:2021tct,Bjerrum-Bohr:2021vuf,Bjerrum-Bohr:2021din,Damgaard:2021ipf,Brandhuber:2021eyq,Bjerrum-Bohr:2021wwt,Bjerrum-Bohr:2022ows,Bern:2021yeh,Bern:2022jvn,Bern:2021dqo,Damgaard:2023ttc,DiVecchia:2023frv,Heissenberg:2023uvo,Adamo:2022ooq,Dlapa:2021npj,Frellesvig:2023bbf,Driesse:2024xad,Klemm:2024wtd,Frellesvig:2024zph,Akpinar:2025bkt,Driesse:2026qiz,Bern:2025wyd,Brandhuber:2025igz,Kim:2025gis,Buonanno:2022pgc,Alaverdian_2025,Bhattacharyya_2025,Bianchi:2025xol,Alessio:2025nzd} and gravitational bremsstrahlung effects \cite{Kosower:2018adc,Brandhuber:2023hhy,Herderschee:2023fxh,Elkhidir:2023dco,Georgoudis:2023lgf,Caron-Huot:2023vxl,Jakobsen:2023hig,Bohnenblust:2023qmy,Bini:2023fiz,Georgoudis:2023eke,Brandhuber:2023hhl,DeAngelis:2023lvf,Adamo:2024oxy,Bini:2024rsy,Alessio:2024wmz,Georgoudis:2024pdz,Brunello:2024ibk,Brandhuber:2024qdn,Bohnenblust:2025gir,Brunello:2025eso} 
in various binary systems. \\[5pt]
When finite-size effects such as tidal deformations, spin, and black-hole absorption become significant, a more fundamental description is required to model black holes as extended objects \cite{Bjerrum-Bohr:2025lpw,Kosmopoulos:2025rfj}. However, at lower post-Minkowskian orders, aspects of this finite-size physics can be adequately described by a local effective-field theory. This is accomplished by working in the context of a multipole expansion and introducing higher-dimensional effective graviton-matter couplings~\cite{Guevara:2017csg,Arkani-Hamed:2017jhn,Guevara:2018wpp,Chung:2018kqs,Guevara:2019fsj,Arkani-Hamed:2019ymq,Aoude:2020onz,Chung:2020rrz,Guevara:2020xjx,Chen:2021kxt,Kosmopoulos:2021zoq,Chiodaroli:2021eug,Bautista:2021wfy,Cangemi:2022bew,Ochirov:2022nqz,Damgaard:2019lfh,Bern:2020buy,Comberiati:2022ldk,Maybee:2019jus,Haddad:2021znf,Chen:2022clh,Menezes:2022tcs,FebresCordero:2022jts,Alessio:2022kwv,Bern:2022kto,Aoude:2022thd,Aoude:2022trd,Alessio:2023kgf,Aoude:2023vdk,Bautista:2023szu,DeAngelis:2023lvf,Brandhuber:2023hhl,Aoude:2023dui,Riva:2022fru,Luna:2023uwd,Chen:2024mmm,Chen:2024bpf,Chen:2024mlx,Ben-Shahar:2025tiz,Gatica:2025uhx,Aoude:2025xxq,Bohnenblust:2024hkw} (see also refs.~\cite{Antonelli:2019ytb,Damour:2020tta,Kalin:2020fhe,Dlapa:2021npj,Dlapa:2021vgp,Dlapa:2022lmu,Jakobsen:2023ndj,Jakobsen:2021smu,Mougiakakos:2021ckm,Jakobsen:2021lvp,Mogull:2020sak,Jakobsen:2021zvh,Vines:2017hyw,Vines:2018gqi,Liu:2021zxr,Jakobsen:2022fcj,Damgaard:2022jem,Bianchi:2023lrg,Gonzo:2024zxo,Hoogeveen:2025tew} for related approaches).\\[5pt]
The heavy-mass effective field theory (HEFT)~\cite{Damgaard:2019lfh,Brandhuber:2021kpo,Brandhuber:2021eyq} is a specialized framework designed for computations in the post-Minkowskian regime. It streamlines the calculation of scattering amplitudes relevant to binary dynamics by universally adopting a limit in which the mass of the black holes is much larger than the momentum they exchange. Like the worldline effective field theory, this approach can also systematically incorporate finite-size effects. For instance, spinning extensions have been developed in \cite{Bjerrum-Bohr:2023iey,Bjerrum-Bohr:2023jau} using physical factorization information and a bootstrap ansatz.\\[5pt]
In this work, we 
consider the gravitational scattering amplitude for a binary system composed of a light and a heavy spinning black hole at the third post-Minkowskian order. We systematically expand to first order in the mass ratio, incorporating spin effects for both bodies.
This approach enables the investigation of phenomena beyond the test-particle (probe) limit, where the backreaction of the light body on the spacetime geometry becomes significant. We construct the required two-loop integrand by assembling tree-level spinning amplitudes derived from the bootstrap approach \cite{Bjerrum-Bohr:2023jau, Bjerrum-Bohr:2023iey}. Following integration using various integration-by-parts techniques \cite{Gehrmann:1999as,Smirnov:2008iw,Larsen:2015ped,Lee:2014ioa}, we obtain the momentum-space amplitude and then perform a Fourier transform to impact-parameter space. This allows us to determine important classical observables such as the angle for the scattering process.\\[5pt]
Our focus in the current presentation is on the aligned-spin configuration only. At the probe limit, zero-self-force limit (0SF), we derive an analytic expression for the eikonal-like phase, expanded in spin parameters. At first order in the self-force expansion (1SF), we provide explicit results up to fifth order in the total spin and quadratic order in the spin of the light black hole. We also separate and analyze the conservative and radiation-reaction contributions, and observe simplifications in their structure. Our 1SF results are new through fifth order in spin. For the conservative sector, they agree with the recent results of ref.~\cite{Akpinar:2025bkt} for the bending angle in the scattering of a Kerr black hole by a Schwarzschild black hole, upon setting one of the black-hole spins to zero, up to quartic order in the Kerr black hole spin.\\[5pt]
In addition, we perform a resummation of the heavy black hole's spin in the probe sector, treating the light black hole with its spin truncated at quadratic order, consistent with \cite{Damgaard:2022jem}. At first self-force and third post-Minkowskian order, we identify two new spin-independent relations among the coefficients of the master integrals. These relations allow us to deduce, for the first time, the fully resummed radiation-reaction contribution valid for arbitrary spins of both black holes. We demonstrate that the resulting eikonal-like phase displays the characteristic ring singularity of the Kerr metric. Additionally, we conjecture that the leading high-energy scattering terms cancel at all orders in spin, as the conservative contribution arises solely from the computed double-box diagram.\\[5pt]
The structure of the paper is as follows. Section \ref{sec:conventions} introduces the kinematics of spinning binary scattering. Section \ref{sec:trees} reviews the heavy-mass effective field theory framework, and presents the spinning tree-level amplitudes that serve as computational building blocks. In Section \ref{sec:0sf}, we compute the eikonal-like phase at zeroth self-force order (probe limit), providing results to high orders in spin and performing a complete spin resummation. Section \ref{sec:1sf} contains our novel findings: the calculation of the eikonal-like phase at first self-force order, including both conservative and radiation-reaction components. Section \ref{sec:resum} examines the resummation of spin effects and their relation to the Kerr singularity. Section \ref{sec:conclusion} concludes with a summary and outlook on future research. Technical details, such as the decomposition of tensors and explicit amplitude coefficients, are provided in the appendices.
\section{Convenient choices of kinematics and conventions}\label{sec:conventions}
We study the scattering between two Kerr black holes with masses $m$ and $M$ and spins $a_1$ and $a_2$. Incoming momenta are denoted by $p_1$ and $p_2$, while outgoing momenta are $p_4$ and $p_3$. The kinematics are summarized in the following diagram:
\begin{equation}
\label{kinematics}
\begin{array}{lr}
\begin{tikzpicture}[scale=12,baseline={([yshift=-1mm]centro.base)}]
\def\x{0}
\def\y{0}
\node at (0+\x,0+\y) (centro) {};
\node at (-3pt+\x,-3pt+\y) (uno) {$p_1$};
\node at (-3pt+\x,3pt+\y) (due) {$p_2$};
\node at (3pt+\x,3pt+\y) (tre) {$p_3$};
\node at (3pt+\x,-3pt+\y) (quat) {$p_4$};
\draw [double] (uno) -- (centro);
\draw [double] (due) -- (centro);
\draw [double] (tre) -- (centro);
\draw [double] (quat) -- (centro);
\draw [->] (-2.8pt+\x,-2pt+\y) -- (-1.8pt+\x,-1pt+\y); 
\draw [->] (1.8pt+\x,-1pt+\y) -- (2.8pt+\x,-2pt+\y) ; 
\draw [->] (-2.8pt+\x,2pt+\y) -- (-1.8pt+\x,1pt+\y)  ; 
\draw [->] (1.8pt+\x,1pt+\y) -- (2.8pt+\x,2pt+\y); 
\node at (0+\x,0+\y) [draw, fill=gray!90!black, circle, inner sep=10pt] {};
\shade [shading=radial] (centro) circle (1.5pt);
\end{tikzpicture}
&\hspace{2cm}
\begin{aligned}
p_1^\mu & \equiv  \bar{p}^\mu_1 + \tfrac{q^\mu}{2}  = (E_1,\vec{p}+\vec{q}/2)\, ,  \\
 p_2^\mu & \equiv  \bar{p}^\mu_2 - \tfrac{q^\mu}{2} = (E_2,-\vec{p}-\vec{q}/2) \, , \\
p_3^\mu & \equiv   \bar{p}^\mu_2 + \tfrac{q^\mu}{2} = (E_3,-\vec{p}+\vec{q}/2) \, ,  \\
p_4^\mu & \equiv  \bar{p}^\mu_1 - \tfrac{q^\mu}{2}  =  (E_4,\vec{p}-\vec{q}/2)\ .
\end{aligned}
\end{array}
\end{equation}
The momentum exchange $q$ is spacelike, and it follows that $\bar p_i \cdot q = 0$. 
The Mandelstam invariants are given by:
$s\equiv (p_1+p_2)^2\text{ and }q^2\equiv (p_1-p_4)^2$.\\[5pt]
We define the barred masses as:
$
 \bar{p}_1^2 \equiv  \bar{m}^2 = m^2-\frac{q^2}{4} \ , \
\bar{p}_2^2 \equiv  \bar{M}^2 = M^2-\frac{q^2}{4} \ ,
$
and introduce the corresponding relativistic velocities: 
$\bar p_1 \equiv  \bar m \bar v_1, \ \bar p_2 \equiv  \bar M \bar v_2, \ \bar v_{i}^2=1.$
We note that their scalar product defines the Lorentz factor:
 $   y\equiv \bar v_1\mdot \bar v_2.$\\[5pt]
In the heavy-mass limit, the spin of each black hole is described by an antisymmetric tensor, $S_i^{\mu\nu}$, which may be expressed in terms of the Pauli-Lubanski pseudovector $a^\mu_i$ as (we refer to ref. \cite{Brandhuber:2023hhl} for further details on the heavy-mass expansion of $S^{\mu\nu}_i$):
\begin{equation}
\label{eq:SpinS}
\begin{aligned}
S_i^{\mu\nu}=\epsilon^{\mu\nu\rho\sigma}\bar p_{i\rho}\,a_{i\sigma},
\end{aligned}
\end{equation}
We note that the spin supplementary condition implies:
$a_1\cdot \bar v_1 =0,\, a_2\cdot \bar v_2 = 0.  $\\[5pt]
Formally, amplitudes in the heavy-mass effective field theory are defined in terms of the barred variables $\bar m$ and $\bar v$. In the classical limit, one may set $\bar m\to m$ and $\bar v \to v$. For simplicity, all amplitudes presented in the following will be written in terms of unbarred variables.\\[5pt]
We work in the regime:
   $     M\gg m,\, |a_1|<|a_2|\leq GM,$
and perform a simultaneous post-Minkowskian expansion and self-force expansion. At the third post-Minkowskian order, we expand in the mass ratio $\mu\equiv m/M$ and retain the terms up to $\terms{\mu}$. For the light, test-like black hole, we keep spin effects up to quadratic order, $\terms{a_1^2}$, while for the heavy, background-like black hole, we keep spin effects to all orders, $\terms{a_2^\infty}$. We note that the post-Minkowskian expansion is organized in powers of $G$ associated with graviton exchange, while the spin dependence is treated as an independent expansion.\\[5pt]
The full amplitude at 3PM is constructed from spinning three-, four-, and five-point tree amplitudes. At this order, the three- and four-point amplitudes are required to all orders in the heavy black hole spin, while the five-point amplitude contributes only up to quadratic order. In the eikonal-like phase, however, only the two-massive-particle-irreducible (2MPI) part of the amplitude contributes.\\[5pt]
We compute the eikonal-like phase $\delta_{\rm H}$, defined in \cite{Brandhuber:2021kpo} as the impact-parameter-space Fourier transform of the 2MPI part of the 3PM amplitude,
\begin{equation}\label{eq:Amp2Phase}
\delta_{\rm H}
= \int \frac{d^{D}q}{(2\pi)^{D-2}}
\frac{e^{ib\cdot q}\delta(q\cdot v_1)\delta(q\cdot v_2)}{4Mm}
\mathcal{M}^{\text{3PM}}_{\rm 2MPI}.
\end{equation}
Throughout this work, we restrict to spin-aligned configurations for which
$ a_1\mdot v_2=a_2\mdot v_1=0, \ a_1\mdot b=a_2\mdot b=0.$ 
For such configurations only the spin magnitudes are relevant, and we introduce the positive dimensionless ratio
$ \xi \equiv \frac{|a_1|}{|a_2|} >0 . $
Aligned configurations satisfy
$ a_1^\mu = \xi\, a_2^\mu , $
while anti-aligned configurations satisfy
$ a_1^\mu = - \xi\, a_2^\mu . $
Equivalently, expressions for anti-aligned configurations may be obtained from those for aligned configurations by the replacement $\xi \to -\xi$.\\[5pt]
We remark that in this setup the scattering occurs in a plane and can therefore be described by a single scattering angle. This angle can be extracted from the eikonal-like phase as follows:
\begin{equation}
    \chi = -\frac{\sqrt{s}}{Mm\sqrt{y^2-1}}\frac{\partial}{\partial{|b|}}\mathrm{Re}[\delta_{\rm H}] \, .
\end{equation}
We work everywhere in $D = 4 - 2\eps$ dimensions and use the mostly minus metric convention $(+,-,-,-)$. For the space-like vectors $q,\, a_1,\, a_2,\, b$ we define :
\begin{equation}
q^2 = -|q|^2, \qquad
a_1^2 = -|a_1|^2, \qquad
a_2^2 = -|a_2|^2, \qquad
b^2 = -|b|^2 . 
\end{equation}
\section{Tree amplitudes}\label{sec:trees}
In this section, we summarize the spinning tree-level amplitudes required for our calculations. We recall that these amplitudes correspond to the $\terms{m^2}$ terms in the inverse-mass expansion of the full amplitude:
\begin{align}\label{eq:heftExpansion}
\begin{tikzpicture}[baseline={([yshift=-0.8ex]current bounding box.center)}]\tikzstyle{every node}=[font=\small]	
\begin{feynman}
    	 \vertex (a) {\(p\)};
    	 \vertex [right=1.5cm of a] (f2)[HV] {$~~~~$ };
    	 \vertex [right=1.5cm of f2] (f3){\(p'\) };
    	 \vertex [below=1.5cm of f2] (gm){$\boldsymbol\cdots$};
    	 \vertex [left=0.8cm of gm] (g2){$k_{1}$};
      \vertex [right=0.8cm of gm] (g20){$k_{n}$};
      \diagram* {
(a) -- [fermion,thick] (f2)-- [fermion,thick] (f3) ,
    	  (g2)--[photon,ultra thick](f2),(g20)--[photon,ultra thick](f2),
    	  };
    \end{feynman}
    \end{tikzpicture}
&=\begin{tikzpicture}[baseline={([yshift=-0.8ex]current bounding box.center)}]\tikzstyle{every node}=[font=\small]	
\begin{feynman}
    	 \vertex (a) {\(p\)};
    	 \vertex [right=1.5cm of a] (f2)[dot]{ };
    	 \vertex [right=1.5cm of f2] (f3)[dot]{ };
      \vertex [right=1.5cm of f3] (f4) [dot] {};
      \vertex [right=1.5cm of f4] (f5) {\(p'\)};
    	 \vertex [below=1.5cm of f2] (gm){$k_{1}$ };
      \vertex [below=1.5cm of f3] (g3) {$k_{2}$};
      \vertex [below=1.5cm of f4] (g4) {$k_{n}$};
      \vertex [right=0.75cm of f2] (att1);
      \vertex [above=0.3cm of att1] (cut20){};
      \vertex [below=0.3cm of att1] (cut21){};
      \vertex [right=0.75cm of f3] (att2);
      \vertex [above=0.3cm of att2] (cut22){};
      \vertex [below=0.3cm of att2] (cut23){};
      \vertex [below=1.5cm of att2] (dots) {$\boldsymbol\cdots$};
    	  \diagram* {
(a) -- [fermion,thick] (f2)-- [thick] (f3)-- [thick](f4)--[fermion, thick](f5),
    	  (gm)--[photon,ultra thick](f2), (g3)--[photon,ultra thick](f3), (g4)--[photon, ultra thick](f4),
          (cut20)--[ red,thick] (cut21), (cut22)--[ red,thick] (cut23)
    	  };
    \end{feynman}  
    \end{tikzpicture}\nn\\
&+
\begin{tikzpicture}[baseline={([yshift=-0.8ex]current bounding box.center)}]\tikzstyle{every node}=[font=\small]	
\begin{feynman}
    	 \vertex (a) {\(p\)};
    	 \vertex [right=1.5cm of a] (f2)[HV]{H};
    	 \vertex [right=3.5cm of f2] (f3)[dot]{ };
      \vertex [right=2.0cm of f2] (f3new)[dot] { };
      \vertex [right=1.5cm of f3] (f4) {\(p'\)};
    	 \vertex [below=1.5cm of f2] (gm){ };
    	 \vertex [left=0.5cm of gm] (g2){$k_{1}$};
      \vertex [right=0.5cm of gm] (g20){$k_{2}$};
      \vertex [below=1.5cm of f3] (g3) {$k_{n}$};
      \vertex [below=1.5cm of f3new] (g3new) {$k_{3}$};
      \vertex [right=1.1cm of f2] (att);
      \vertex [above=0.3cm of att] (cut20){};
      \vertex [below=0.3cm of att] (cut21){};
      \vertex [right=2.75cm of f2] (att2);
      \vertex [above=0.3cm of att2] (cut30){};
      \vertex [below=0.3cm of att2] (cut31){};
      \vertex [below=1.5cm of att2] (dots) {$\boldsymbol\cdots$};
    	  \diagram* {
(a) -- [fermion,thick] (f2)-- [thick] (f3)-- [fermion, thick](f4),
    	  (g2)--[photon,ultra thick](f2), (g20)--[photon,ultra thick](f2), (g3)--[photon,ultra thick](f3),(g3new)--[photon,ultra thick](f3new), (cut20)--[ red,thick] (cut21),(cut30)--[ red,thick] (cut31)
    	  };
    \end{feynman}  
    \end{tikzpicture} + \text{Perms.}\nn\\
    &+\begin{tikzpicture}[baseline={([yshift=-0.8ex]current bounding box.center)}]\tikzstyle{every node}=[font=\small]	
\begin{feynman}
    	 \vertex (a) {\(p\)};
    	 \vertex [right=1.5cm of a] (f2)[HV]{H};
    	 \vertex [right=3.5cm of f2] (f3)[dot]{ };
      \vertex [right=2.0cm of f2] (f3new)[dot] { };
      \vertex [right=1.5cm of f3] (f4) {\(p'\)};
    	 \vertex [below=1.5cm of f2] (gm){$k_{2}$ };
    	 \vertex [left=0.75cm of gm] (g2){$k_{1}$};
      \vertex [right=0.75cm of gm] (g20){$k_{3}$};
      \vertex [below=1.5cm of f3] (g3) {$k_{n}$};
      \vertex [below=1.5cm of f3new] (g3new) {$k_{4}$};
      \vertex [right=1.1cm of f2] (att);
      \vertex [above=0.3cm of att] (cut20){};
      \vertex [below=0.3cm of att] (cut21){};
      \vertex [right=2.75cm of f2] (att2);
      \vertex [above=0.3cm of att2] (cut30){};
      \vertex [below=0.3cm of att2] (cut31){};
      \vertex [below=1.5cm of att2] (dots) {$\boldsymbol\cdots$};
    	  \diagram* {
(a) -- [fermion,thick] (f2)-- [thick] (f3)-- [fermion, thick](f4),
    	  (g2)--[photon,ultra thick](f2), (g20)--[photon,ultra thick](f2),(gm)--[photon,ultra thick](f2), (g3)--[photon,ultra thick](f3),(g3new)--[photon,ultra thick](f3new), (cut20)--[ red,thick] (cut21),(cut30)--[ red,thick] (cut31)
    	  };
    \end{feynman}  
    \end{tikzpicture} + \text{Perms.}\nn\\
&+\hspace{4cm}\boldsymbol\vdots\nn\\
&+\begin{tikzpicture}[baseline={([yshift=-0.8ex]current bounding box.center)}]\tikzstyle{every node}=[font=\small]	
\begin{feynman}
    	 \vertex (a) {\(p\)};
    	 \vertex [right=1.5cm of a] (f2)[HV]{H};
    	 \vertex [right=1.5cm of f2] (f3){\(p'\) };
    	 \vertex [below=1.5cm of f2] (gm){$\boldsymbol\cdots$};
    	 \vertex [left=0.8cm of gm] (g2){$k_{1}$};
    	  \vertex [right=0.8cm of gm] (g20){$k_{n}$};
    	  \diagram* {
(a) -- [fermion,thick] (f2)-- [fermion,thick] (f3) ,
    	  (g2)--[photon,ultra thick](f2),(g20)--[photon,ultra thick](f2),
    	  };
    \end{feynman}
    \end{tikzpicture} +(\text{derivative of}~ \delta +\text{quantum}).
\end{align} 
The ellipsis in eq.~\eqref{eq:heftExpansion} indicates terms in which additional gravitons attach directly to the heavy vertex $H$. The red lines denote delta functions of the scalar product between the velocity and the soft graviton momentum, $\delta(2m\, v\!\cdot\! k_1)$, which arise from expanding massive propagators in the heavy-mass limit (see ref. \cite{Brandhuber:2021eyq} for details.). These are also referred to in the literature as velocity or massive cuts \cite{Bjerrum-Bohr:2021vuf,Bjerrum-Bohr:2021din,Bjerrum-Bohr:2021wwt,Bjerrum-Bohr:2022blt,Bjerrum-Bohr:2022ows}. The blobs labeled $H$ represent amplitudes in the heavy-mass limit, and the plain grey blob represents the full amplitude. We omit contributions involving derivatives of delta functions, as well as quantum terms, since they
do not give classical contributions. The amplitudes used in our analysis correspond to the terms without delta-function insertions in the above expansion.
\subsection{The spinning three-point amplitude}
The minimally coupled three-point amplitude with two massive legs, of mass $m$ and spin $a$, and one graviton in the heavy-mass limit, is given in \cite{Vines:2017hyw,Arkani-Hamed:2017jhn,
Guevara:2018wpp, Chung:2018kqs, Chen:2021kxt} as:
\begin{align}
\label{eq:3pointV1}
\mathcal{M}_3\left(1, v\right)=\begin{tikzpicture}[baseline={([yshift=-0.8ex]current bounding box.center)}]\tikzstyle{every node}=[font=\small]	
\begin{feynman}
    	 \vertex (a) {\(v, a\)};
    	 \vertex [right=1.5cm of a] (f2)[dot]{ };
    	 \vertex [right=1.5cm of f2] (f3){\(\)};
    	 \vertex [below=1.5cm of f2] (gm){$k_{1}$};
    	  \diagram* {
(a) -- [fermion,thick] (f2)-- [fermion,thick] (f3),
    	 (gm)--[photon,ultra thick](f2)
    	  };
\end{feynman}  
\end{tikzpicture} =m^2 \left(v \mdot\vareps_1\right)^2\exp{\left(i\frac{k_1\mdot S\mdot\vareps_1}{m v\mdot \vareps_1}\right)},
\end{align}
where $S^{\mu\nu}$ is defined in eq.~\eqref{eq:SpinS} and we use the abbreviation $\varepsilon_1\equiv \varepsilon\left(k_1\right)$. Eq.~\eqref{eq:3pointV1}, can be rewritten in the more convenient form~\cite{Bjerrum-Bohr:2023jau}:
\begin{equation}\label{eq:spinning3point}
\mathcal{M}_3(1,v) = m(v\mdot \varepsilon_1)(w_1\mdot\varepsilon_1),
\end{equation}
with $w^\mu_i\equiv  w(k_i)^\mu$ defined as \begin{equation}
\begin{aligned}
    w_j^{\mu} &= \cosh\left(k_j\mdot a\right)m v^{\mu} + i \left(G_1(k_j\mdot a)\right)\left(k_j\mdot S\right)^{\mu},\ \ \ 
     G_1(x) &= \frac{\sinh x}{x}.
\end{aligned}
\end{equation}
\subsection{Compton amplitude}
We start from the spinning Compton amplitude presented in \cite{Bjerrum-Bohr:2023iey,Bjerrum-Bohr:2023jau} and expand it in the small-spin parameter up to $\mathcal{O}(a^5)$. The resulting amplitude can be written schematically as:
\begin{align}\label{amplitude structure}
    \cM_4(1,2, v)&=-\sum_{i=1}^{(5)}{\npre^{(i)}_a(1,2, v)\,\npre_0(1,2, v)\over 
    k^2_{12}}+\sum_{i=3}^{5}\cM^{(i)}_{4,r}(1,2,v)+\sum_{i=4}^{5}\cM^{(i)}_{4,c}(1,2,v).
\end{align}
Here, the first term represents the double-copy contribution, while $\cM_{4,r}$ and $\cM_{4,c}$ denote the remainder and contact terms, which first appear at $\mathcal{O}(a^3)$ and $\mathcal{O}(a^4)$, respectively and which do not have a double-copy representation. \\[5pt]
To perform the small-spin expansion, we decompose the spin vector as $a^\mu = |a| \hat a^\mu$, with $\hat a \cdot \hat a = -1$, and expand the numerators $\npre_a$, $\cM_{4,r}$, and $\cM_{4,c}$ of \cite{Bjerrum-Bohr:2023jau} around $|a| \ll 1$. We use the shorthand $k^\mu_{ij\cdots n} = k^\mu_i + k^\mu_j + \cdots + k^\mu_n$. \\[5pt]
The resulting Compton amplitude agrees with independent computations up to at least $\mathcal{O}(a^8)$ for the leading contribution in the parameter $z=\sqrt{-a\mdot a} v\mdot k_1$, as shown in \cite{Chen:2024mmm}. Corrections associated with this parameter characterize composite-particle effects of the spinning black hole, which start at $\mathcal{O}(a^5)$. We will not consider those here. \\[5pt]
For the double copy part, the related spinning numerators $\npre^{(i)}_{a}$ are:
\begin{align}
\npre^{(0)}_a(1,2, v)&=-\frac{v\mdot F_1\mdot F_2\mdot v}{k_1\mdot v}\nn,\\
\npre^{(1)}_a(1,2, v)&=i \text{tr}\left(S\mdot F_2\mdot F_1\right)+\frac{i k_1\mdot S\mdot F_1\mdot F_2\mdot v}{k_1\mdot v}+\frac{i k_2\mdot S\mdot F_2\mdot F_1\mdot v}{k_1\mdot v}\nn,\\
\npre^{(2)}_a(1,2, v)&=-\left(a\mdot F_1\mdot k_2\right) a\mdot F_2\mdot v+a\mdot F_2\mdot k_1 a\mdot F_1\mdot v+a\mdot F_1\mdot F_2\mdot a k_1\mdot v\nn\\
&-\frac{\left(\left(a\mdot k_1\right){}^2+\left(a\mdot k_2\right){}^2\right) v\mdot F_1\mdot F_2\mdot v}{2 k_1\mdot v}-\frac{k_1\mdot S\mdot F_1\mdot F_2\mdot S\mdot k_2}{k_1\mdot v}\nn,\\
\npre^{(3)}_a(1,2, v)&=\frac{1}{6} i \left(a\mdot k_{1,2}\right){}^2 \text{tr}\left(S\mdot F_2\mdot F_1\right)+\frac{i \left(\left(a\mdot k_1\right){}^2+3 \left(a\mdot k_2\right){}^2\right) k_1\mdot S\mdot F_1\mdot F_2\mdot v}{6 k_1\mdot v}\nn\\
&+\frac{i \left(3 \left(a\mdot k_1\right){}^2+\left(a\mdot k_2\right){}^2\right) k_2\mdot S\mdot F_2\mdot F_1\mdot v}{6 k_1\mdot v}\nn\\
&+\frac{1}{3} i \left(a\mdot k_1-a\mdot k_2\right) (a\mdot F_1\mdot F_2\mdot S\mdot k_2+ a\mdot F_2\mdot F_1\mdot S\mdot k_1)\nn,\\
\npre^{(4)}_a(1,2, v)&=-\frac{\left(\left(a\mdot k_1\right){}^2+\left(a\mdot k_2\right){}^2\right)}{6 }\Big({k_1\mdot S\mdot F_1\mdot F_2\mdot S\mdot k_2\over k_1\mdot v}+a\mdot F_1\mdot k_2 a\mdot F_2\mdot v\nn\\
&~~~~~~~~~~~~~~~~~~~~~~~~~~~~~~~~~~-a\mdot F_2\mdot k_1 a\mdot F_1\mdot v-a\mdot F_1\mdot F_2\mdot a  k_1\mdot v\Big) \\
&  -\frac{\left(\left(a\mdot k_1\right){}^4+6 \left(a\mdot k_2\right){}^2 \left(a\mdot k_1\right){}^2+\left(a\mdot k_2\right){}^4\right) }{24 k_1\mdot v}v\mdot F_1\mdot F_2\mdot v\nn,\\
\npre^{(5)}_a(1,2, v)&=\frac{1}{120} i \left(a\mdot k_{1,2}\right){}^4 \text{tr}\left(S\mdot F_2\mdot F_1\right)\nn\\
&+\frac{i \left(\left(a\mdot k_1\right){}^4+10 \left(a\mdot k_2\right){}^2 \left(a\mdot k_1\right){}^2+5 \left(a\mdot k_2\right){}^4\right) k_1\mdot S\mdot F_1\mdot F_2\mdot v}{120 k_1\mdot v}\nn\\
&+\frac{i \left(5 \left(a\mdot k_1\right){}^4+10 \left(a\mdot k_2\right){}^2 \left(a\mdot k_1\right){}^2+\left(a\mdot k_2\right){}^4\right) k_2\mdot S\mdot F_2\mdot F_1\mdot v}{120 k_1\mdot v}\nn\\
&+\frac{1}{30} i \left(a\mdot k_1-a\mdot k_2\right) \left(\left(a\mdot k_1\right){}^2+\left(a\mdot k_2\right){}^2\right)\Big( a\mdot F_1\mdot F_2\mdot S\mdot k_2+a\mdot F_2\mdot F_1\mdot S\mdot k_1\Big).\nn
\end{align}
One can observe that all expressions are manifestly gauge invariant and involve only physical heavy-mass propagators. The scalar products of the field strength tensors, $F_i^{\mu\nu} = k_i^{\mu}\varepsilon_i^{\nu} -k_i^{\nu}\varepsilon_i^{\mu}$, reappear across different spin orders due to expanding the entire functions of $a\mdot k_1, a\mdot k_2$ introduced in \cite{Bjerrum-Bohr:2023iey,Bjerrum-Bohr:2023jau}. \\[5pt]
For the other contributions we have: 
\begin{align}
\cM_{4,r}^{(3)}(1,2, v)	&=-\frac{i \left(a\mdot F_1\mdot F_2\mdot v+a\mdot F_2\mdot F_1\mdot v\right) \left(\text{tr}\left(S\mdot F_1\right) a\mdot F_2\mdot v-\text{tr}\left(S\mdot F_2\right) a\mdot F_1\mdot v\right)}{12 k_1\mdot v},\\
\cM_{4,r}^{(4)}(1,2, v)&=\frac{1}{12\, k_1\mdot v}
\Big[
(a\mdot k_1 - a\mdot k_2)\,(a\mdot F_1\mdot F_2\mdot a+a^2 v\mdot F_1\mdot F_2\mdot v)\,
\bigl(
a\mdot F_1\mdot F_2\mdot v
+a\mdot F_2\mdot F_1\mdot v
\bigr)
\Big] \nn,\\
\cM_{4,c}^{(4)}(1,2, v)&=\frac{1}{6}
\Big[
 a\mdot F_1\mdot F_2\mdot a\,(2
a\mdot F_1\mdot v\,
a\mdot F_2\mdot v
+ a^2\,
v\mdot F_1\mdot F_2\mdot v\,
)
- a^2\,
a\mdot F_1\mdot F_2\mdot v\,
a\mdot F_2\mdot F_1\mdot v
\Big] \nn ,\\
\cM_{4,r}^{(5)}(1,2, v)&=
-\frac{i}{60\, k_1\mdot v}
\Big[
\bigl(
((a\mdot k_1)^2- a\mdot k_1\, a\mdot k_2+(a\mdot k_2)^2)
(
a\mdot F_1\mdot F_2\mdot v
+a\mdot F_2\mdot F_1\mdot v
)
\nonumber\\
&
~~~~~~~~~~~~~\times\bigl(
\tr(S\mdot F_1)\, a\mdot F_2\mdot v
-\tr(S\mdot F_2)\, a\mdot F_1\mdot v
\bigr)
\Big]\nn,\\
 \cM_{4,c}^{(5)}(1,2, v)&=
\frac{i}{20\, }
\Big[
\bigl(
(a\mdot k_1 - a\mdot k_2)\, a\mdot F_1\mdot F_2\mdot a\,
\bigr)
\bigl(
\tr(S\mdot F_1)\, a\mdot F_2\mdot v
-\tr(S\mdot F_2)\, a\mdot F_1\mdot v
\bigr)
\Big].\nn
\end{align}
\subsection{Five-point amplitude up to quadratic order}
Similarly to the Compton amplitude, the only contribution to the tree-level spinning five-point amplitude up to $\terms{a^2}$ is given by the double-copy terms:\vskip-25pt
\begin{equation}
\begin{aligned}
\label{eq:5pointHEFT}
\mathcal{M}_{5}\left(1,2,3,v\right) =\begin{tikzpicture}[baseline={([yshift=-0.8ex]current bounding box.center)}]\tikzstyle{every node}=[font=\small]	
\begin{feynman}
    	 \vertex (a) {\(v, a\)};
    	 \vertex [right=1.5cm of a] (f2)[HV]{H};
    	 \vertex [right=1.5cm of f2] (f3){\(\) };
    	 \vertex [below=1.5cm of f2] (gm){$k_{2}$};
    	 \vertex [left=0.8cm of gm] (g2){$k_{1}$};
    	  \vertex [right=0.8cm of gm] (g20){$k_{3}$};
    	  \diagram* {
(a) -- [fermion,thick] (f2)-- [fermion,thick] (f3) ,
    	  (g2)--[photon,ultra thick](f2),(gm)--[photon,ultra thick](f2),(g20)--[photon,ultra thick](f2),
    	  };
    \end{feynman}
    \end{tikzpicture} &= {\npre_a([1,[2,3]],v)\,\npre_0([1,[2,3]],v)\over k_{23}^2\,k_{123}^2}\\[-15pt]
& +{\npre_a([[1,2],3],v)\,\npre_0([[1,2],3],v)\over k_{12}^2\,k_{123}^2}\\ 
& +{\npre_a([[1,3],2],v)\,\npre_0([[1,3],2],v)\over k_{13}^2\,k_{123}^2}.
\end{aligned}    
\end{equation}\\[-15pt]
Here $\npre_0$ denotes the five-point scalar numerator which is constructed from kinematic Hopf algebra \cite{Brandhuber:2021bsf,Chen:2022nei,Chen:2024gkj,Fu:2025jpp}:
\begin{equation}\label{eq:heftSpin05Point}
\begin{aligned}
    \npre_0\left([[1,2],3],v\right) &= m \Big(\frac{v\mdot F_1\mdot F_2\mdot F_3\mdot v}{v\mdot k_1}- \frac{v\mdot F_1\mdot F_2 \mdot v \, k_{12}\mdot F_3\mdot v}{v\mdot k_1 v\mdot k_{12}}  - \frac{v\mdot F_1\mdot F_3\mdot v \, k_1\mdot F_2\mdot v}{v\mdot k_1 v\mdot k_{13}}\Big).
\end{aligned}
\end{equation}
$\npre_a$ is the spinning numerator presented in \cite{Bjerrum-Bohr:2023jau}. We expand $\npre_a$ around $|a|\ll1$, and up to $\terms{a^2}$:
\begin{equation}
    \npre_a\left([[1,2],3],v \right) = \npre_a^{(0)}\left([[1,2],3],v\right) + \npre_{a}^{(1)}\left([[1,2],3],v\right) + \npre_{a}^{(2)}\left([[1,2],3],v\right),
\end{equation}
where $\npre_a^{(0)}=\npre_{0}$ is given by eq.~\eqref{eq:heftSpin05Point}, and $\npre_{a}^{(1)}$ and $\npre_{a}^{(2)}$ are given by:

\begin{align}
    &\npre_{a}^{(1)}\left([[1,2],3],v\right) =i m\Big[ \frac{ \text{tr}\left(S\mdot F_2 \mdot F_1\right) k_{12}\mdot F_3\mdot v}{k_{12}\mdot v}+\frac{\text{tr}\left(S\mdot F_3\mdot F_1\right) k_1\mdot F_2\mdot v}{k_{13}\mdot v}+\frac{ \text{tr}\left(S\mdot F_3\mdot F_2\right) k_2\mdot F_1\mdot v}{k_1\mdot v}\nn\\
    &-\frac{ v\mdot F_1\mdot F_2\mdot v k_3\mdot S\mdot F_3\mdot k_{12}}{k_1\mdot v k_{1,2}\mdot v}+\frac{ k_{12}\mdot F_3\mdot v k_1\mdot S\mdot F_1\mdot F_2\mdot v}{k_1\mdot v k_{12}\mdot v}+\frac{ k_1\mdot F_2\mdot v k_1\mdot S\mdot F_1\mdot F_3\mdot v}{k_1\mdot v k_{13}\mdot v}\nn\\
    &+\frac{ k_{12}\mdot F_3\mdot v k_2\mdot S\mdot F_2\mdot F_1\mdot v}{k_1\mdot v k_{12}\mdot v}+\frac{ k_1\mdot F_2\mdot v k_3\mdot S\mdot F_3\mdot F_1\mdot v}{k_1\mdot v k_{13}\mdot v}-\frac{i v\mdot F_1\mdot F_3\mdot v k_1\mdot F_2\mdot S\mdot k_2}{k_1\mdot v k_{13}\mdot v}\nn\\
    &+\frac{ k_3\mdot S\mdot F_3\mdot F_2\mdot F_1\mdot v}{k_1\mdot v}-\frac{ k_1\mdot S\mdot F_1\mdot F_2\mdot F_3\mdot v}{k_1\mdot v}- \text{tr}\left(S \mdot F_2\mdot F_1\mdot F_3\right)+ \text{tr}\left(S\mdot F_3\mdot F_2\mdot F_1\right) \Big],
\end{align}

and

\begin{align}
&\npre_{a}^{(2)}\left([[1,2],3],v\right) =\frac{\bigl((a\mdot k_1)^2+(a\mdot k_2)^2+(a\mdot k_3)^2\bigr)}{2}\npre_{a}^{(0)}\left([[1,2],3]\right)\nn\\
&-\frac{\text{tr}\left(S\mdot F_2\mdot F_1\right) k_3\mdot S\mdot F_3\mdot k_{12}}{k_{12}\mdot v}-\frac{\text{tr}\left(S\mdot F_3\mdot F_2\right) k_1\mdot S\mdot F_1\mdot k_2}{k_1\mdot v}-\frac{\text{tr}\left(S\mdot F_3\mdot F_1\right) k_1\mdot F_2\mdot S\mdot k_2}{k_{13}\cdot v}\nn\\
&+\frac{k_{12}\mdot F_3\mdot v \left(a\mdot k_1 a\mdot F_2\mdot F_1\mdot v-a\mdot F_1\mdot k_2 a\mdot F_2\mdot v\right)}{k_{12}\mdot v}+\frac{k_1\mdot F_2\mdot v \left(a\mdot k_1 a\mdot F_3\mdot F_1\mdot v-a\mdot F_1\mdot k_3 a\mdot F_3\mdot v\right)}{k_{13}\mdot v}\nn\\
&-\frac{k_3\mdot S\mdot F_3\mdot k_{12} \left(k_1\mdot S\mdot F_1\mdot F_2\mdot v+k_2\mdot S\mdot F_2\mdot F_1\mdot v\right)}{k_1\mdot v k_{12}\mdot v}-\frac{k_1\mdot F_2\mdot S\mdot k_2 \left(k_1\mdot S\mdot F_1\mdot F_3\mdot v+k_3\mdot S\mdot F_3\mdot F_1\mdot v\right)}{k_1\mdot v k_{13}\mdot v}\nn\\
&+\frac{k_2\mdot F_1\mdot v \left(a\mdot k_2 a\mdot F_3\mdot F_2\mdot v-a\mdot F_2\mdot k_3 a\mdot F_3\mdot v\right)}{k_1\mdot v}-a\mdot k_{12} \left(a\mdot F_3\mdot F_1\mdot F_2\mdot v-a\mdot F_3\mdot F_2\mdot F_1\mdot v\right)\nn\\
&+\left(a\mdot F_1\mdot F_2\mdot k_3-a\mdot F_2\mdot F_1\mdot k_3\right) a\mdot F_3\mdot v+a\mdot F_1\mdot k_2 \left(a\mdot F_2\mdot F_3\mdot v-a\mdot F_3\mdot F_2\mdot v\right)\nn\\
&+a\mdot F_2\mdot k_1 a\mdot F_3\mdot F_1\mdot v-a\mdot k_1 a\mdot F_2\mdot F_1\mdot F_3\mdot v.
\end{align}

\section{Binary scattering at zero self-force order}\label{sec:0sf}
At zeroth self-force and third post-Minkowskian (PM) order, we have the diagram:
\begin{align}\label{fig:TwoLoopProbe}
&\mathcal{M}^{3\text{PM}}_{0\text{SF}}\left(q,a_1,a_2,v_1,v_2\right)=\begin{tikzpicture}[baseline={([yshift=-0.8ex]current bounding box.center)}]\tikzstyle{every node}=[font=\small]	
\begin{feynman}
    	 \vertex (a) {\(v_1, a_1\)}; 
         \vertex [right=3cm of a] (v1) [HV]{H};	 
    	 \vertex [right=3cm of v1] (b){};
    	 \vertex [above=2.0cm of a](c){$v_2, a_2$};   \vertex [right=1.5cm of c] (u1) [dot]{};
         \vertex [right=1.5cm of u1] (u2) [dot]{};
         \vertex [right=1.5cm of u2] (u3) [dot]{};
    	  \vertex [right=1.5cm of u3](d){};
    	  \vertex [above=1.0cm of a] (cutL);
    	  \vertex [right=6.0cm of cutL] (cutR);
    	   \vertex [right=0.75cm of u1] (cut1);
    	  \vertex [above=0.3cm of cut1] (cut1u);
    	  \vertex [below=0.3cm of cut1] (cut1b);
    	  \vertex [right=0.75cm of u2] (cut2);
    	  \vertex [above=0.3cm of cut2] (cut2u);
    	  \vertex [below=0.3cm of cut2] (cut2b);
    	  \diagram* {
(a) -- [fermion,thick] (v1)-- [fermion,thick] (b),
    	  (u1)--[photon,ultra thick,momentum'=\(\ell_1\)](v1), (u2)-- [photon,ultra thick,momentum'=\(\ell_2\)] (v1),(u3)-- [photon,ultra thick,momentum=\(\ell_3\)] (v1), (c) -- [fermion,thick] (u1)-- [fermion,thick] (u2)-- [fermion,thick] (u3)-- [fermion,thick] (d), (cutL)--[dashed, red,thick] (cutR), (cut1u)--[red,thick] (cut1b),(cut2u)--[red,thick] (cut2b),
    	  };
    \end{feynman}  
    \end{tikzpicture}\nn\\
  &=\frac{1}{3!}\frac{(32\pi G_N)^3}{(4\pi)^D}\pi^2\int {d^{D}\ell_1\over \pi^{D/2}} {d^{D}\ell_3\over \pi^{D/2}} \delta(-M v_2\mdot \ell_1)\delta(-M v_2\mdot \ell_3 )\times \nn\\
  &\sum_{\rm polarizations} {\mathcal{M}_5(\ell_1, \ell_2, \ell_3,v_1) 
  \mathcal{M}_3(-\ell_1,v_2)\mathcal{M}_3(-\ell_2,v_2)\mathcal{M}_3(-\ell_3,v_2)\over \ell_1^2 \ell_2^2 \ell_3^2 }.
\end{align}
The spinning five-point and three-point tree amplitudes are given by eqs.~\eqref{eq:5pointHEFT}  and \eqref{eq:spinning3point}. We eliminate the momentum $\ell_2^\mu$ in favor of $\ell_1^\mu$, $\ell_3^\mu$, and $q^\mu$ using momentum conservation, $q^\mu = \ell_1^\mu + \ell_2^\mu + \ell_3^\mu$. Following the discussion in \cite{Kosmopoulos:2020pcd}, we then perform the polarization sum using the completeness relation in harmonic gauge:
\begin{align}\label{gluingf}
\sum_{\rm polarizations}\varepsilon^{*\mu_a} \varepsilon^{*\nu_a}\varepsilon^{\mu_b} \varepsilon^{\nu_b}={1\over 2}\Big[ \eta^{\mu_a \mu_b}\eta^{\nu_a \nu_b}+\eta^{\mu_a \nu_b}\eta^{\nu_a \mu_b}-{2\over (D-2)}\eta^{\mu_a\nu_a}\eta^{\mu_b\nu_b}\Big].
 \end{align}
The integrand eq.~\eqref{fig:TwoLoopProbe}, is expanded in powers of the spin $a_2$. As in the one-loop analysis of \cite{Chen:2024mmm}, the integrand decomposes into two parts. The first is a scalar part, in which the spin dependence enters only through scalar products with the loop momenta and external vectors, $a_2\cdot p_x$. The second is a tensor part, which contains contractions of the loop momenta $\ell_i$ and external vectors, with the spin tensors $S_1^{\mu\nu}$ and $S_2^{\mu\nu}$, defined in eq.~\eqref{eq:SpinS}:
\begin{equation}
\begin{aligned}
&\ell_i\mdot S_1\mdot p_x,  & \ell_i\mdot S_2\mdot p_x,
\end{aligned}
\end{equation}
where $p_x^\mu\in \{q^\mu, v_1^\mu,v_2^\mu,\ell_{1}^\mu,\ell_{3}^\mu\}$ and $\ell_i\in\{\ell_1, \ell_3\}$.\\[5pt]
In principle, the spin tensors can be decomposed into a basis built from the external vectors, as shown in the Appendix~\ref{ap:epsdecomp}. However, in the present case, it is more practical to decompose the underlying Levi–Civita tensor $\epsilon\left(a_2,\mu,\nu,\rho\right)$, with $\mu, \, \nu\text{ and }\rho$ being free indices. This yields:
 \begin{equation}
 \label{eq:epsDecomp}
     \begin{aligned}
         \epsilon\left(a_2,\mu,\nu,\rho\right) =\frac{q\cdot S_2\cdot v_1\left(\left(q\cdot a_2\right)
   v_{1}^{[\mu} v_{2}^{\nu}a_2^{\rho]} - a_2^2 v_{1}^{[\mu} v_{2}^{\nu}q^{\rho]} \right)}{M\left(y^2-1\right) \left(\left(q\cdot a_2\right){}^2-a_2^2
   q^2\right)}.
     \end{aligned}
 \end{equation}
The notation $a^{[\mu} b^{\nu} c^{\rho]}$ denotes the anti-symmetrization of the indices 
$\mu,\,\nu,\,\rho$. \\[5pt]
After contracting eq.~\eqref{eq:epsDecomp} with $\ell_1$ and $\ell_3$, the integral structures  are reduced to:
 \begin{equation}
 \label{eq:loopints}
         \int \lm \frac{\delta\left(\ell_1\cdot v_2\right)\delta\left(\ell_3\cdot v_2\right)\left(a_2\cdot\ell_1\right)^{\lambda_1}\left(a_2\cdot\ell_3\right)^{\lambda_2}\left(\ell_1\cdot \ell_3\right)^{\lambda_{3}}\left(\ell_{1}\cdot q\right)^{\lambda_{4}}\left(\ell_{3}\cdot q\right)^{\lambda_{5}}}{\ell_1^2\ell_3^2\left(q-\ell_1-\ell_3\right)^2\left(\ell_1\cdot v\right)^{\lambda_6} D_1^{\lambda_7}D_2^2},\\
 \end{equation}
 where $D_1$ can be $\ell_3 \cdot v_1$ or $(\ell_1 + \ell_3)\cdot v_1$, and $D_2$ denotes massless propagators such as $(q - \ell_3)^2$, $(q - \ell_1)^2$ or $(\ell_1 + \ell_3)^2$, depending on the channel.\\[5pt]
The integral in eq.~\eqref{eq:loopints} is evaluated using integration-by-parts (IBP) identities with LiteRed \cite{Lee:2012cn,Lee:2013mka}. At each order in spin, the loop integrals reduce to the  double triangle fan-integral:
\begin{align}\label{eq:doubleTriAmp}
	&\mathcal{M}^{3\text{PM}}_{0\text{SF}}\left(q,a_1,a_2,v_1,v_2\right)= \mathcal{I}_{\rm fan} \Big[\Big(\frac{\alpha ^4 \left(64 y^6-104 y^4+44 y^2-3\right)}{256 \left(y^2-1\right)^2}\nn\\
	&+\!\frac{\alpha ^2 \left(128 y^6\!-\!216 y^4\!+\!96 y^2\!-\!7\right)}{64 \left(y^2-1\right)^2}+\frac{64 y^6\!-\!120 y^4\!+\!60 y^2\!-\!5}{32 \left(y^2-1\right)^2} + \terms{\alpha^5} \Big)\!+\!\frac{q\cdot S_2\cdot v_1}{M}\nn\\
	&\times\!\Big(\frac{i \alpha ^4 y \left(112 y^4\!-\!124 y^2\!+\!27\right) }{1920 \left(y^2-1\right)^2}\!+\!\frac{i \alpha ^2 y \left(80 y^4\!-\!92 y^2\!+\!21\right) }{96 \left(y^2-1\right)^2}\!+\!\frac{3 i y \left(16 y^4\!-\!20 y^2\!+\!5\right) }{16 \left(y^2-1\right)^2}\!+\!\terms{\alpha^ 5}\Big)\nn\\
	&+\!\xi\Big(\frac{\alpha ^6 \left(224 y^6-352 y^4+142 y^2-9\right)}{3840 \left(y^2-1\right)^2}+\frac{\alpha ^4 \left(160 y^6-256 y^4+106 y^2-7\right)}{192 \left(y^2-1\right)^2}\nn\\
	&+\!\frac{\alpha ^2 \left(96 y^6\!-\!160 y^4\!+\!70 y^2\!-\!5\right)}{32 \left(y^2-1\right)^2}\! +\! \terms{\alpha^7}\Big)\!+\!\xi \frac{q\cdot S_2\cdot v_1}{M} \nn\\
	&\times\!\Big(\!\frac{i \alpha ^4 y \left(48 y^4\!-\!52 y^2\!+\!11\right)}{192 \left(y^2-1\right)^2}\!+\!\frac{i \alpha ^2 y \left(8 y^4\!-\!9 y^2\!+\!2\right)}{4 \left(y^2-1\right)^2}\!+\!\frac{i y \left(16 y^4\!-\!20 y^2\!+\!5\right)}{8 \left(y^2-1\right)^2}\! +\! \terms{\alpha^5}\Big)\nn\\
	&+\!\xi^2\Big(\frac{\alpha ^6 \left(192 y^6-296 y^4+116 y^2-7\right)}{1536 \left(y^2-1\right)^2}+\frac{\alpha ^4 \left(128 y^6-200 y^4+80 y^2-5\right)}{128 \left(y^2-1\right)^2}\nn\\
	&+\!\frac{\alpha ^2 \left(64 y^6-104 y^4+44 y^2-3\right)}{64 \left(y^2-1\right)^2}+\terms{\alpha^7}\Big)+\xi^2 \frac{q\cdot S_2\cdot v_1}{M}\nn\\
	&\times\!\Big(\frac{i \alpha ^6 y \left(112\! y^4-\!116 y^2\!+\!23\right)}{3840 \left(y^2-1\right)^2}\!+\!\frac{i \alpha ^4 y \left(80 y^4\!-\!84 y^2\!+\!17\right)}{192 \left(y^2-1\right)^2}\!+\!\frac{i \alpha ^2 y \left(48 y^4\!-\!52 y^2\!+\!11\right)}{32 \left(y^2-1\right)^2}\!+\!\terms{\alpha^7}\Big),
\end{align}
where $\alpha^2\equiv  (a_2\mdot q)^2-a_2^2 q^2$ and 
\begin{equation}
     \label{eq:MI}
    \mathcal{I}_{\rm fan} \equiv  \int \lm \frac{\delta\left(\ell_1\cdot v_2\right)\delta\left(\ell_3\cdot v_2\right)}{\ell_1^2\ell_3^2\left(q-\ell_1-\ell_3\right)^2} = \frac{4\pi^2}{\eps}\left(-\frac{q^2}{2}\right)^{-2\eps}\, .
 \end{equation} \\[5pt]
The eikonal-like phase is obtained directly by eq.~\eqref{eq:Amp2Phase}, using:
\begin{equation}
    \label{eq:IPSFourier}
    \begin{aligned}
   & \int \qm \frac{e^{ib\cdot q}\delta\left(q\cdot v_1\right)\delta\left(q\cdot v_2\right)}{4Mm}(-q^2)^{\nu}q^{\mu_1}\ldots q^{\mu_n} \\&= (-i)^{n}{\partial\over \partial b_{\mu_1}}\ldots {\partial\over \partial b_{\mu_n}} \int \qm \frac{e^{ib\cdot q}\delta\left(q\cdot v_2\right)\delta\left(q\cdot v_2\right)}{4Mm}(-q^2)^{\nu}\\
&= (-i)^{n}{\partial\over \partial b_{\mu_1}}\ldots {\partial\over \partial b_{\mu_n}}\left(\frac{\pi^{-(D-2)/2}}{4}\frac{2^{D-2+2\nu}}{\sqrt{y_1^2 - 1}\left(-b^2\right)^{(D-2)/2 + \nu}}\frac{\Gamma\left(\frac{D-2}{2}+\nu\right)}{\Gamma\left(-\nu\right)}\right).
    \end{aligned}
\end{equation} 
The four vectors $q^\mu$ (or the derivatives $ \partial/\partial b_\mu$) are contracted with either the spin vector $a^\mu_2$, or $(v_1\cdot S_2)^{\mu}$, or both. After all integrations, the remaining scalar product $v_1\cdot S_2\cdot b$ can be reduced further as we show in eq.~\eqref{eq:LCres}. \\[5pt]
The resulting eikonal-like phase depends only on three dimensionless parameters $y,\xi$, and $\beta\equiv  {|a_2|\over |b|}$. All dependence on dimensionful parameters is contained in an overall prefactor, $G_N^3 M^3 m\over |b|^2$. The eikonal-like phase is denoted as $\delta^{(j, k)}_{\rm H, 0SF}$, where the first index indicates the spin order of the heavy black hole and the second index indicates the spin order of the light black hole. The total spin order ({\it i.e.} $\beta$-order) is $j+k$. At order $\mathcal{O}(\xi^0)$, the resulting eikonal-like phase up to $\terms{\beta^5}$ is:
\begin{equation}\label{eq:probeAtxi0}
\scalebox{1.08}{$
\begin{array}{c||c|c|c}
G_N^3 M^3 m{1\over |b|^2} & i=0 & i=1 & i=2 \\ \hline\hline
\delta_{\rm H, 0SF}^{(2i,0)}
& \frac{64 y^6-120 y^4+60 y^2-5}{3  (y^2-1)^{5/2}}
&
 \frac{\beta^2 (128 y^6-216 y^4+96 y^2-7)}{ (y^2-1)^{5/2}}
&
 \frac{5 \beta^4 (64 y^6-104 y^4+44 y^2-3)}{ (y^2-1)^{5/2}}
\\[0.5em] \hline
\delta_{\rm H, 0SF}^{(2i+1,0)}
&
 \frac{4 \beta y (16 y^4-20 y^2+5)}{ (y^2-1)^2}
&
\frac{8 \beta^3 y (80 y^4-92 y^2+21)}{3  (y^2-1)^2}
&
 \frac{4 \beta^5 y (112 y^4-124 y^2+27)}{ (y^2-1)^2}
\\ \hline
\end{array}$}
\end{equation}
At $\terms{\xi^1}$, the phase up to $\terms{\beta^6}$ is shown below:
\begin{equation}\label{eq:probeAtxi1}
\scalebox{1.08}{$
\begin{array}{c||c|c|c}
G_N^3 M^3 m{\xi \over |b|^2} & i=0 & i=1 & i=2 \\ \hline\hline
\delta_{\rm H, 0SF}^{(2i,1)}
&-\frac{8 \beta y (16 y^4-20 y^2+5)}{3  (y^2-1)^2}
& -\frac{64 \beta^3 y (8 y^4-9 y^2+2)}{ (y^2-1)^2}
& -\frac{40 \beta^5 y (48 y^4-52 y^2+11)}{ (y^2-1)^2}
\\[0.5em] \hline
\delta_{\rm H, 0SF}^{(2i+1,1)}
& \frac{2 \beta^2 (96 y^6-160 y^4+70 y^2-5)}{ (y^2-1)^{5/2}}
& \frac{20 \beta^4 (160 y^6-256 y^4+106 y^2-7)}{3  (y^2-1)^{5/2}}
& \frac{14 \beta^6 (224 y^6-352 y^4+142 y^2-9)}{ (y^2-1)^{5/2}}
\\ \hline
\end{array}
$}
\end{equation}
At $\terms{\xi^2}$, the phase up to $\terms{\beta^7}$ is shown below:
\begin{equation}\label{eq:probeAtxi2}
\scalebox{1.08}{$
\begin{array}{c||c|c|c}
G_N^3 M^3 m {\xi^2\over |b|^2}  & i=0 & i=1 & i=2 \\ \hline\hline
\delta_{\rm H, 0SF}^{(2i,2)}
& \frac{\beta^2 (64 y^6-104 y^4+44 y^2-3)}{ (y^2-1)^{5/2}}
& \frac{10 \beta^4 (128 y^6-200 y^4+80 y^2-5)}{ (y^2-1)^{5/2}}
& \frac{35 \beta^6 \!\left(4 y^2 (48 y^4-74 y^2+29)-7\right)}{ (y^2-1)^{5/2}}
\\[0.5em] \hline
\delta_{\rm H, 0SF}^{(2i+1,2)}
& -\frac{8 \beta^3 y (48 y^4-52 y^2+11)}{ (y^2-1)^2}
& -\frac{40 \beta^5 y (80 y^4-84 y^2+17)}{ (y^2-1)^2}
& -\frac{112 \beta^7 y (112 y^4-116 y^2+23)}{ (y^2-1)^2}
\\ \hline
\end{array}$}
\end{equation}
The results  are in agreement with \cite{Damgaard:2022jem} and with \cite{Akpinar:2025bkt} up to $\terms{\beta^4,\xi^0}$. 
\section{Binary scattering at first self-force order}\label{sec:1sf}
At the first self-force order, the relevant scattering amplitude is obtained by combining contributions from the four diagrams shown below.
\begin{align}\label{eq:amp1sf}
\mathcal{M}^{3\text{PM}}_{1\text{SF}}\left(q,a_2,a_1,v_2,v_1\right)&=\mathcal{M}^{3\text{PM}}_{1\text{SF},\left(\ell_1,\ell_2,\ell_3\right)}\!\cup\!\mathcal{M}^{3\text{PM}}_{1\text{SF},\left(\ell_2,\ell_4,\ell_5\right)} \!\cup\!\mathcal{M}^{3\text{PM}}_{1\text{SF},\left(\ell_1,\ell_2,\ell_4\right)}\!\cup\!\mathcal{M}^{3\text{PM}}_{1\text{SF},\left(\ell_2,\ell_3,\ell_5\right)},\nn\\
&=
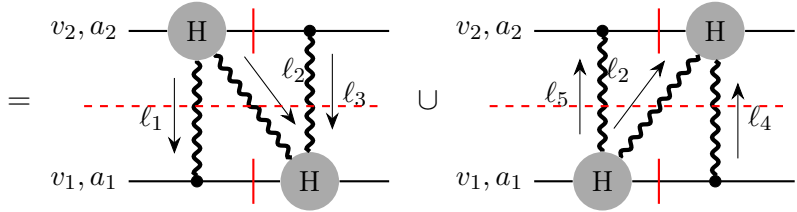
\begin{tikzpicture}[baseline={([yshift=-0.2ex]current bounding box.center)}]\tikzstyle{every node}=[font=\small]	
\begin{feynman}
    	 \vertex (p1) {\(v_1,a_1\)};
    	 \vertex [right=1.5cm of p1] (b1) [dot]{};
    	  \vertex [right=1.5cm of b1] (b2) [HV]{H};
    	 \vertex [right=1.2cm of b2] (p4){};
    	 \vertex [above=2.0cm of p1](p2){$v_2,a_2$};
    	 \vertex [right=1.5cm of p2] (u1) [HV]{H};
    	 \vertex [right=1.5cm of u1] (u2) [dot]{};
    	  \vertex [right=1.2cm of u2](p3){};
    	  \vertex [above=1.cm of p1] (cutL);
    	  \vertex [right=3.9cm of cutL] (cutR);
    	  \vertex [right=0.75cm of u1] (cut1);
    	  \vertex [above=0.3cm of cut1] (cut1u);
    	  \vertex [below=0.3cm of cut1] (cut1b);
    	   \vertex [right=0.75cm of b1] (cutb1);
    	  \vertex [above=0.3cm of cutb1] (cutb1u);
    	  \vertex [below=0.3cm of cutb1] (cutb1b);
    	  \diagram* {
(p1) -- [thick] (b1)-- [thick] (b2) -- [thick] (p4),
    	  (u1)--[photon,ultra thick,momentum'=$\ell_1$](b1), (u2)-- [photon,ultra thick,momentum=$\ell_3$] (b2),(u1)-- [photon,ultra thick,momentum=$\ell_2$] (b2), (p2) -- [thick] (u1)-- [thick] (u2)-- [thick] (p3), (cutL)--[dashed, red,thick] (cutR), (cut1u)--[ red,thick] (cut1b),(cutb1u)--[red,thick] (cutb1b)
    	  };
    \end{feynman}  
    \end{tikzpicture}  \cup 
     \begin{tikzpicture}[baseline={([yshift=-0.2ex]current bounding box.center)}]\tikzstyle{every node}=[font=\small]	
\begin{feynman}
    	 \vertex (p1) {\(v_1,a_1\)};
    	 \vertex [right=1.5cm of p1] (b1) [HV]{H};
    	  \vertex [right=1.5cm of b1] (b2) [dot]{};
    	 \vertex [right=1.2cm of b2] (p4){};
    	 \vertex [above=2.0cm of p1](p2){$v_2,a_2$};
    	 \vertex [right=1.5cm of p2] (u1) [dot]{};
    	 \vertex [right=1.5cm of u1] (u2) [HV]{H};
    	  \vertex [right=1.2cm of u2](p3){};
    	  \vertex [above=1.cm of p1] (cutL);
    	  \vertex [right=3.9cm of cutL] (cutR);
    	  \vertex [right=0.75cm of u1] (cut1);
    	  \vertex [above=0.3cm of cut1] (cut1u);
    	  \vertex [below=0.3cm of cut1] (cut1b);
    	   \vertex [right=0.75cm of b1] (cutb1);
    	  \vertex [above=0.3cm of cutb1] (cutb1u);
    	  \vertex [below=0.3cm of cutb1] (cutb1b);
    	  \diagram* {
(p1) -- [thick] (b1)-- [thick] (b2) -- [thick] (p4),
    	  (b1)--[photon,ultra thick,momentum=$\ell_5$](u1), (b2)-- [photon,ultra thick,momentum'=$\ell_4$] (u2),(b1)-- [photon,ultra thick,momentum=$\ell_2$] (u2), (p2) -- [thick] (u1)-- [thick] (u2)-- [thick] (p3), (cutL)--[dashed, red,thick] (cutR), (cut1u)--[ red,thick] (cut1b),(cutb1u)--[red,thick] (cutb1b)
    	  };
    \end{feynman}  
    \end{tikzpicture}\nn\\
&\cup \begin{tikzpicture}[baseline={([yshift=-0.2ex]current bounding box.center)}]\tikzstyle{every node}=[font=\small]	
\begin{feynman}
    	 \vertex (p1) {\(v_1,a_1\)};
    	 \vertex [right=1.5cm of p1] (b1) [dot]{};
    	  \vertex [right=1.5cm of b1] (b2) [dot]{};
    	 \vertex [right=1.2cm of b2] (p4){};
    	 \vertex [above=2.0cm of p1](p2){$v_2,a_2$};
    	 \vertex [right=1.5cm of p2] (u1) [HV]{H};
    	 \vertex [right=1.5cm of u1] (u2) [HV]{H};
    	  \vertex [right=1.2cm of u2](p3){};
    	  \vertex [above=1.cm of p1] (cutL);
    	  \vertex [right=3.9cm of cutL] (cutR);
    	  \vertex [right=0.75cm of u1] (cut1);
    	  \vertex [above=0.3cm of cut1] (cut1u);
    	  \vertex [below=0.3cm of cut1] (cut1b);
    	   \vertex [right=0.75cm of b1] (cutb1);
    	  \vertex [above=0.3cm of cutb1] (cutb1u);
    	  \vertex [below=0.3cm of cutb1] (cutb1b);
    	  \diagram* {(u1)-- [photon,ultra thick,out=-45,in=-135,looseness=0.5,min distance=0.7cm,momentum'=$\ell_2$] (u2),(p1) -- [thick] (b1)-- [thick] (b2) -- [thick] (p4),
    	  (u1)--[photon,ultra thick,momentum'=$\ell_1$](b1), (b2)-- [photon,ultra thick,momentum'=$\ell_4$] (u2), (p2) -- [thick] (u1)-- [thick] (u2)-- [thick] (p3), (cutL)--[dashed, red,thick] (cutR), (cut1u)--[ red,thick] (cut1b),(cutb1u)--[red,thick] (cutb1b),(cutb1u)--[dashed,red,thick] (cut1b)
    	  };
    \end{feynman}  
    \end{tikzpicture} \cup 
    \begin{tikzpicture}[baseline={([yshift=-0.2ex]current bounding box.center)}]\tikzstyle{every node}=[font=\small]	
\begin{feynman}
    	 \vertex (p1) {\(v_1,a_1\)};
    	 \vertex [right=1.5cm of p1] (b1) [HV]{H};
    	  \vertex [right=1.5cm of b1] (b2) [HV]{H};
    	 \vertex [right=1.2cm of b2] (p4){};
    	 \vertex [above=2.0cm of p1](p2){$v_2,a_2$};
    	 \vertex [right=1.5cm of p2] (u1) [dot]{};
    	 \vertex [right=1.5cm of u1] (u2) [dot]{};
    	  \vertex [right=1.2cm of u2](p3){};
    	  \vertex [above=1.cm of p1] (cutL);
    	  \vertex [right=3.9cm of cutL] (cutR);
    	  \vertex [right=0.75cm of u1] (cut1);
    	  \vertex [above=0.3cm of cut1] (cut1u);
    	  \vertex [below=0.3cm of cut1] (cut1b);
    	   \vertex [right=0.75cm of b1] (cutb1);
    	  \vertex [above=0.3cm of cutb1] (cutb1u);
    	  \vertex [below=0.3cm of cutb1] (cutb1b);
    	  \diagram* {(b1)-- [photon,ultra thick,out=45,in=135,looseness=0.5,min distance=0.7cm,momentum=$\ell_2$] (b2),(p1) -- [thick] (b1)-- [thick] (b2) -- [thick] (p4),
    	  (b1)--[photon,ultra thick,momentum=$\ell_5$](u1), (u2)-- [photon,ultra thick,momentum=$\ell_3$] (b2), (p2) -- [thick] (u1)-- [thick] (u2)-- [thick] (p3), (cutL)--[dashed, red,thick] (cutR), (cut1u)--[ red,thick] (cut1b),(cutb1u)--[red,thick] (cutb1b),(cutb1u)--[dashed,red,thick] (cut1b)
    	  };
    \end{feynman}  
    \end{tikzpicture}.
  \end{align}
The $\cup$ operation denotes that the diagrams are combined such that the contribution from each master integral is counted only once. Furthermore, we define $q=\ell_1+\ell_2+\ell_3$ as the transfer momentum. The internal momenta $\ell_4,\ell_5$ can be expressed in terms of $\ell_1,\, \ell_2$, and $q$: \begin{align}
   \ell_5=-\ell_1-\ell_2,&&\ell_4=-\ell_2-\ell_3=-q+\ell_1.
\end{align}
For each diagram, we construct the integrand by gluing the relevant tree-level amplitudes. For instance, the integrand resulting from the first graph in eq.~\eqref{eq:amp1sf} will be:  
\begin{equation}
    \begin{aligned}
&\mathcal{M}^{3\text{PM}}_{1\text{SF},\left(\ell_1,\ell_2,\ell_3\right)}=\frac{1}{2!}\frac{(32\pi G_N)^3}{(4\pi)^{D-2}}\int {d^{D}\ell_1\over \pi^{D/2}} {d^{D}\ell_3\over \pi^{D/2}} \delta(-m v_1\mdot \ell_1)\delta(-M v_2\mdot \ell_3 )\times\\&\sum_{\rm polarizations}\frac{\cM_4(\ell_1,\ell_2, v_2)\cM_3(\ell_3, v_2)\cM_4(\ell_2,\ell_3, v_1)\cM_3(\ell_1, v_1)}{ \ell_1^2 \ell_2^2 \ell_3^2},
    \end{aligned}
\end{equation}
where $\cM_3$ is the three point amplitude given in eq.~\eqref{eq:3pointV1}, $\cM_4$ is the four point amplitude given in eq.~\eqref{amplitude structure}, and we perform the polarization sums using eq.~\eqref{gluingf}. 
As in the probe-limit case, the integrands contain tensor structures of the form 
\begin{align}\label{cases}
    (\ell_i\cdot (S_X)\cdot p_x),
\end{align}
where $p_x$ can be any vector, $p_x\in \{\ell_{i}, a_2, q, v_1, v_2\}$, and $\ell_i\in\{\ell_1,\ell_2,\ell_3\}$, denote the loop momenta.\\[5pt]
In contrast to the probe limit, where only the spin tensors $S_1$ and $S_2$ appeared, the present case involves general rank-2 tensors $S_X$, constructed from the available vectors and spin tensors.\\[5pt]
To resolve these contractions, we decompose $S_X$ into a basis of the external vectors $\{v_1, v_2,a_2, q\}$, following the general procedure outlined in Appendix~\ref{ap:epsdecomp}.\\[5pt]
The integrals in the amplitude can be written in terms of the following family:
\begin{equation}
\begin{aligned}
G_{i_1 , i_2 , i_3 , i_4 , i_5 , i_6 , i_7 , i_8 , i_9 , i_{10}, i_{11}}&\!\!=\!\int\! {d^{D}\ell_1\over \pi^{D/2}} {d^{D}\ell_3\over \pi^{D/2}}\frac{\delta^{i_2-1}(-v_2\mdot \ell_3 )\delta^{i_3-1}(-v_1\mdot \ell_1)(a_2\mdot \ell_1)^{-{i_{10}}}(a_2\mdot \ell_2)^{-{i_{11}}}}{(v_2\mdot \ell_1)^{i_1}(v_1\mdot \ell_3)^{i_4}(\ell_1^2)^{i_5}(\ell_2^2)^{i_6}(\ell_3^2)^{i_7}(\ell_4^2)^{i_8}(\ell_5^2)^{i_9}},
\end{aligned}
\end{equation}
After performing IBP reduction, the 6918 distinct integrals appearing in the amplitude are expressed in terms of a set of seven master integrals. We choose these master integrals to form a canonical basis, so that the amplitude becomes: 
\begin{equation}\label{redamp}
    \begin{aligned}
   \mathcal{M}^{3\text{PM}}_{1\text{SF}}&= C_1 \mathcal{I}'_{1}+C_3 \mathcal{I}'_{3}+C_5 \mathcal{I}'_{5}&&\text{conservative related}\\
  &+{1\over 2}(C_2-C_4) (\mathcal{I}'_{2}-\mathcal{I}'_{4})+C_7 \mathcal{I}'_{7} &&\text{radiation-reaction related}
\\
   &+{1\over 2}(C_2+C_4) (\mathcal{I}'_{2}+ \mathcal{I}'_{4})+C_6 \mathcal{I}'_{6} &&\text{\shortstack{real part in 4-dim, \\ irrelevant for scattering angle.}}
\end{aligned}
\end{equation}
The canonical master integrals $\mathcal{I}'_i$ are defined through the integral family \\$G_{i_1 , i_2 , i_3 , i_4 , i_5 , i_6 , i_7 , i_8 , i_9 , i_{10}, i_{11}}$: 
\begin{align}
G_{0,1,1,0,0,1,0,1,1,0,0}
&= \frac{\mathcal{I}'_{1}}{16 \sqrt{y^2-1}\,\epsilon ^4},
&\qquad
G_{0,1,1,0,0,2,0,1,1,0,0}
&= \frac{-\mathcal{I}'_{2}}{4\epsilon ^3 q^2\sqrt{y^2-1}\,}, \nn \\[4pt]
G_{0,2,2,0,0,1,0,1,1,0,0}
&= \frac{\sqrt{y^2-1}\mathcal{I}'_{3}-y\,\mathcal{I}'_{2}}{2 \epsilon ^2 q^2\sqrt{y^2-1} },
&\qquad
G_{0,1,1,0,1,1,0,1,0,0,0}
&= \frac{\sqrt{y^2-1}\,\mathcal{I}'_{4}}
        {4 \epsilon ^2 \left(8 \epsilon ^2-6 \epsilon +1\right)}, \nn \\[4pt]
G_{0,1,1,0,1,1,1,1,1,0,0}
&= \frac{\mathcal{I}'_{5}}
        {4 \epsilon ^4 q^4 \sqrt{y^2-1}\,},
&\qquad
G_{-1,1,1,-1,1,1,1,1,1,0,0}
&= \frac{\mathcal{I}'_{6}}
        {16 q^2\,\epsilon ^4}, \nn \\[4pt]
G_{1,1,1,1,0,1,0,1,1,0,0}
&= \frac{\mathcal{I}'_{7}}
        {8\epsilon ^4 q^2(y^2-1)\,},
\end{align}
and their topologies are summarized below:
\begin{align}
	\mathcal{I}'_1:\begin{tikzpicture}[baseline={([yshift=-0.8ex]current bounding box.center)}]\tikzstyle{every node}=[font=\small]	
\begin{feynman}
    	 \vertex (p1) {\(p_1\)};
    	 \vertex [right=0.8cm of p1] (b1) [dot]{};
    	 \vertex [right=0.5cm of b1] (b2)[dot]{};
    	 \vertex [right=0.8cm of b2] (p4){$p_4$};
    	 \vertex [above=0.8cm of p1](p2){$p_2$};
    	 \vertex [right=0.8cm of p2] (u1) [dot]{};
    	 \vertex [right=0.5cm of u1] (u2) [dot]{};
    	  \vertex [above=0.8cm of p4](p3){$p_3$};
    	  \diagram* {
(p1) -- [ultra thick] (b1)-- [ultra thick] (b2)-- [ultra thick] (p4),
    	   (b1)-- [thick] (u1),(b2) --  [thick] (u2), (u1)--[thick](b2),(p2) -- [ultra thick] (u1)-- [ultra thick] (u2)-- [ultra thick] (p3),
    	  };
    \end{feynman}  
\end{tikzpicture},&&\mathcal{I}'_2:\begin{tikzpicture}[baseline={([yshift=-0.8ex]current bounding box.center)}]\tikzstyle{every node}=[font=\small]	
\begin{feynman}
    	 \vertex (p1) {\(p_1\)};
    	 \vertex [right=0.8cm of p1] (b1) [dot]{};
    	 \vertex [right=0.5cm of b1] (b2)[dot]{};
    	 \vertex [right=0.8cm of b2] (p4){$p_4$};
    	  \vertex [right=0.25cm of b1] (la)[]{};
    	   \vertex [above=0.4cm of la] (g2)[]{$\circ$};
    	 \vertex [above=0.8cm of p1](p2){$p_2$};
    	 \vertex [right=0.8cm of p2] (u1) [dot]{};
    	 \vertex [right=0.5cm of u1] (u2) [dot]{};
    	  \vertex [above=0.8cm of p4](p3){$p_3$};
    	   \vertex [right=0.25cm of u1] (lb)[]{};
    	  \diagram* {
(p1) -- [ultra thick] (b1)-- [ultra thick] (b2)-- [ultra thick] (p4),
    	   (b1)-- [thick] (u1),(b2) --  [thick] (u2), (u1)--[thick](b2),(p2) -- [ultra thick] (u1)-- [ultra thick] (u2)-- [ultra thick] (p3),
    	  };
    \end{feynman}  
    \end{tikzpicture},&&
	\mathcal{I}'_3:\begin{tikzpicture}[baseline={([yshift=-0.8ex]current bounding box.center)}]\tikzstyle{every node}=[font=\small]	
\begin{feynman}
    	 \vertex (p1) {\(p_1\)};
    	 \vertex [right=0.8cm of p1] (b1) [dot]{};
    	 \vertex [right=0.5cm of b1] (b2)[dot]{};
    	 \vertex [right=0.8cm of b2] (p4){$p_4$};
    	  \vertex [right=0.25cm of b1] (la)[]{$\circ$};
    	 \vertex [above=0.8cm of p1](p2){$p_2$};
    	 \vertex [right=0.8cm of p2] (u1) [dot]{};
    	 \vertex [right=0.5cm of u1] (u2) [dot]{};
    	  \vertex [above=0.8cm of p4](p3){$p_3$};
    	   \vertex [right=0.25cm of u1] (lb)[]{$\circ$};
    	  \diagram* {
(p1) -- [ultra thick] (b1)-- [ultra thick] (b2)-- [ultra thick] (p4),
    	   (b1)-- [thick] (u1),(b2) --  [thick] (u2), (u1)--[thick](b2),(p2) -- [ultra thick] (u1)-- [ultra thick] (u2)-- [ultra thick] (p3),
    	  };
    \end{feynman}  
  \end{tikzpicture},&&\mathcal{I}'_4:\begin{tikzpicture}[baseline={([yshift=-0.8ex]current bounding box.center)}]\tikzstyle{every node}=[font=\small]	
\begin{feynman}
    	 \vertex (p1) {\(p_1\)};
    	 \vertex [right=0.8cm of p1] (b1) [dot]{};
    	 \vertex [right=0.5cm of b1] (b2)[dot]{};
    	 \vertex [right=0.8cm of b2] (p4){$p_4$};
    	 \vertex [above=0.8cm of p1](p2){$p_2$};
    	 \vertex [right=0.8cm of p2] (u1) [dot]{};
    	 \vertex [right=0.5cm of u1] (u2) [dot]{};
    	  \vertex [above=0.8cm of p4](p3){$p_3$};
    	  \diagram* {
(p1) -- [ultra thick] (b1)-- [ultra thick] (b2)-- [ultra thick] (p4),
    	   (b1)-- [thick] (u1),(b1) --  [thick,out=65,in=115,looseness=0.5,min distance=0.4cm] (b2), (u2)--[thick](b2),(p2) -- [ultra thick] (u1)-- [ultra thick] (u2)-- [ultra thick] (p3),
    	  };
    \end{feynman}  
\end{tikzpicture},\nn\\
	\mathcal{I}'_5:\begin{tikzpicture}[baseline={([yshift=-0.8ex]current bounding box.center)}]\tikzstyle{every node}=[font=\small]	
\begin{feynman}
    	 \vertex (p1) {\(p_1\)};
    	 \vertex [right=0.8cm of p1] (b1) [dot]{};
    	 \vertex [right=0.5cm of b1] (b2)[dot]{};
    	 \vertex [right=0.8cm of b2] (p4){$p_4$};
    	  \vertex [right=0.25cm of b1] (la)[]{};
    	   \vertex [above=0.4cm of b1] (g1)[dot]{};
    	    \vertex [above=0.4cm of b2] (g2)[dot]{};
    	 \vertex [above=0.8cm of p1](p2){$p_2$};
    	 \vertex [right=0.8cm of p2] (u1) [dot]{};
    	 \vertex [right=0.5cm of u1] (u2) [dot]{};
    	  \vertex [above=0.8cm of p4](p3){$p_3$};
    	   \vertex [right=0.25cm of u1] (lb)[]{};
    	  \diagram* {
(p1) -- [ultra thick] (b1)-- [ultra thick] (b2)-- [ultra thick] (p4),
    	   (b1)-- [thick] (u1),(b2) --  [thick] (u2), (g1)--[thick](g2),(p2) -- [ultra thick] (u1)-- [ultra thick] (u2)-- [ultra thick] (p3),
    	  };
    \end{feynman}  
    \end{tikzpicture},&&
	\mathcal{I}'_6:\begin{tikzpicture}[baseline={([yshift=-0.8ex]current bounding box.center)}]\tikzstyle{every node}=[font=\small]	
\begin{feynman}
    	 \vertex (p1) {\(p_1\)};
    	 \vertex [right=0.8cm of p1] (b1) [dot]{};
    	 \vertex [right=0.5cm of b1] (b2)[dot]{};
    	 \vertex [right=0.8cm of b2] (p4){$p_4$};
    	  \vertex [right=0.25cm of b1] (la)[]{};
    	     \vertex [above=0.4cm of b2] (n1)[]{$~~~~~~~~v_1\mdot \ell_3$};
    	   \vertex [above=0.4cm of b1] (g1)[dot]{};
    	    \vertex [above=0.4cm of b2] (g2)[dot]{};
    	 \vertex [above=0.8cm of p1](p2){$p_2$};
    	 \vertex [right=0.8cm of p2] (u1) [dot]{};
    	 \vertex [right=0.5cm of u1] (u2) [dot]{};
    	  \vertex [above=0.8cm of p4](p3){$p_3$};
    	   \vertex [right=0.25cm of u1] (lb)[]{};
    	    \vertex [above=0.4cm of b1] (n2)[]{$v_2\mdot \ell_1~~~~~~~~$};
    	  \diagram* {
(p1) -- [ultra thick] (b1)-- [ultra thick] (b2)-- [ultra thick] (p4),
    	   (b1)-- [thick] (u1),(b2) --  [thick] (u2), (g1)--[thick](g2),(p2) -- [ultra thick] (u1)-- [ultra thick] (u2)-- [ultra thick] (p3),
    	  };
    \end{feynman}  
    \end{tikzpicture},&&
    \mathcal{I}'_7:\begin{tikzpicture}[baseline={([yshift=-0.8ex]current bounding box.center)}]\tikzstyle{every node}=[font=\small]	
\begin{feynman}
    	 \vertex (p1) {\(p_1\)};
    	 \vertex [right=0.8cm of p1] (b1) [dot]{};
    	 \vertex [right=0.5cm of b1] (b2)[dot]{};
    	 \vertex [right=0.8cm of b2] (p4){$p_4$};
    	  \vertex [right=0.25cm of b1] (la)[]{};
    	   \vertex [right=0.25cm of b1] (b3)[dot]{};
    	 \vertex [above=0.8cm of p1](p2){$p_2$};
    	 \vertex [right=0.8cm of p2] (u1) [dot]{};
    	 \vertex [right=0.5cm of u1] (u2) [dot]{};
    	  \vertex [above=0.8cm of p4](p3){$p_3$};
    	   \vertex [right=0.25cm of u1] (lb)[]{};
    	    \vertex [right=0.25cm of u1] (u3)[dot]{};
    	  \diagram* {
(p1) -- [ultra thick] (b1)-- [ultra thick] (b2)-- [ultra thick] (p4),
    	   (b1)-- [thick] (u1),(b2) --  [thick] (u2), (b3)--[thick](u3),(p2) -- [ultra thick] (u1)-- [ultra thick] (u2)-- [ultra thick] (p3),
    	  };
    \end{feynman}  
    \end{tikzpicture}.&&
    \end{align}
The canonical basis chosen here coincides with that of \cite{Brandhuber:2021eyq}. We list the values of the integrals as follows:
\begin{align}
\mathcal{I}'_{1}&=2 \epsilon ^3\, \arccosh(y) \mathcal{I}_{3,2}'+\mathcal{O}(\epsilon ^4,i) \ , &
	\mathcal{I}'_{2}&=\epsilon ^2 \mathcal{I}_{2,2}'+\mathcal{O}(\epsilon ^3,i) \ , \nn\\
	\mathcal{I}'_{3}&=\epsilon ^2 \mathcal{I}_{3,2}'+\mathcal{O}(\epsilon ^3,i) \ , &
	\mathcal{I}'_{4}&=-\epsilon ^2 \mathcal{I}_{2,2}'+\mathcal{O}(\epsilon ^3,i) \ , \nn\\
	\mathcal{I}'_{5}&=-\epsilon ^3\, \arccosh(y) \mathcal{I}_{3,2}'+\mathcal{O}(\epsilon ^4,i) \ , &
	\mathcal{I}'_{6}&=\mathcal{O}(\epsilon ^4) \ , \nn\\
	\mathcal{I}'_{7}&=4 \epsilon ^3\, \arccosh(y) \mathcal{I}_{2,2}'+\mathcal{O}(\epsilon ^4,i) \ .
\end{align}
In these equations $\mathcal{O}(\epsilon ^j,i)$ denotes the imaginary part or terms of order greater than or equal to $\epsilon ^j$. The boundary values of these integrals are given at the static limit\footnote{In the differential equation approach of \cite{Brandhuber:2021eyq} these values are used to fix the integration constants.} $y\to1$:
\begin{align}\label{eq:bdvalue}
	\mathcal{I}'_{2,2}&=16\pi^2 \left(-{q^2\over 2}\right)^{-2\epsilon} +\mathcal{O}(i)\, , &	\mathcal{I}'_{3,2}&=32\pi^2 \left(-{q^2\over 2}\right)^{-2\epsilon}.
\end{align}
We note that $\mathcal{I}'_{3,2}$ is associated with the conservative region, while $\mathcal{I}'_{2,2}$ originates from the radiation-reaction region; see \cite{Brandhuber:2021eyq} for a detailed discussion. \\[5pt]
The master integral coefficients $C_i$, depend on the scalar products $a_2^2$, $q^2$, $a_2 \cdot q$ and $v_1\cdot S_2\cdot q$, and can be found in Appendix~\ref{app:amp1sf}. It is worth noting that the expression for the coefficient $C_7$ is remarkably simple:
\begin{align}\label{eq:c7}
	C_7&= {64 G_N^3 M^3m^3\over\pi}\Big[\frac{\left(2 y^2-3\right) y \left(\left(6 \xi ^2+4 \xi +1\right) \left(a_2\cdot q\right){}^4+12 (\xi +1)^2 \left(a_2\cdot q\right){}^2+24\right)}{49152 \left(y^2-1\right)^2 \epsilon ^4}\nn\\
	&+\frac{\left(2 y^2-3\right) \left(8 y^4-8 y^2+1\right) y \left(\alpha ^4 \left(6 \xi ^2+4 \xi +1\right)+12 \alpha ^2 (\xi +1)^2+24\right)}{49152 \left(y^2-1\right)^2 \epsilon ^4} \, \nn\\
	&+\frac{i \left(4 y^4-8 y^2+3\right) y^2  v_1\cdot S_2\cdot q}{61440M \left(y^2-1\right)^2 \epsilon ^4} \nn\\
	&~~~~~~~~~~~~~\times \left(\alpha ^4 \left(10 \xi ^2+5 \xi +1\right)+20 \alpha ^2 \left(3 \xi ^2+3 \xi +1\right)+120 (\xi +1)\right)\Big].
\end{align}
Moreover, we observe that the coefficients $C_2-C_4, C_5$ can be related to $C_7$ through the simple expressions:
\begin{align}\label{eq:c5c24}
	C_5&=\frac{4 \sqrt{y^2-1}  \left(y^2-1\right)}{y \left(2 y^2-3\right)} C_7,\nn\\
	C_2-C_4&=-\frac{4 \sqrt{y^2-1}  \left(5 y^2-8\right) \epsilon }{3y(2 y^2-3)} C_7\, .
\end{align}
The first relation ensures the cancellation of the leading divergence in the limit $y\to\infty$ between the conservative and radiation-reaction contributions to the bending angle at fifth order in spin.\\[5pt]
At first sight, the coefficients $C_2$ and $C_4$ receive contributions from contact terms in the tree-level Compton amplitude, whereas $C_5$ arises solely from the double-copy part. The relations \eqref{eq:c5c24} therefore imply that the contact-term contributions to $C_2$ and $C_4$ cancel non-trivially in the combination $C_2-C_4$.\\[5pt]
Similarly to the probe-limit case for aligned-spin scattering, the eikonal-like phase depends on the dimensionless parameters $\beta,  \xi$ and $y$. In the Appendix \ref{app:chi}, we present our results for the real part of these phases $\delta^{(j,i)}_{\rm H, 1SF}, j\in [0,5-i], i\in [0,2]$, where $j$ and $i$ denote the power of $a_2$ and $\xi$ respectively.\\[5pt]
One can directly verify that, in the high-energy limit (\( y \to \infty \)), the leading behavior of both the conservative and radiation-reaction parts scales as \( y^2 \ln y \). However, in the total contribution to the scattering, these leading terms cancel between the conservative and radiation-reaction parts at arbitrary spin order. This result agrees with the original cancellation proposed in~\cite{DiVecchia:2020ymx} for spinless scattering, as well as with~\cite{Jakobsen:2022zsx} up to quadratic order in spin and~\cite{Akpinar:2025bkt,Haddad:2025cmw} up to quartic order in spin.
\section{Spin resummation and the Kerr singularity at third post-Minkowskian order}\label{sec:resum}
At the third post-Minkowskian order, finite-size effects arising from spin are closely related to the well-known singularity of the Kerr spacetime. However, this connection is not visible at any finite order in the spin expansion, as each individual spin contribution is regular; the Kerr singularity arises once one sums the infinite tower of spin-induced multipole moments of the black hole. We have therefore analyzed our results to identify patterns at arbitrary spin order, organizing the amplitude as a power series in spin that facilitates resummation. We will see that this allows us to reveal the expected non-trivial full spin dependence of the heavy black hole.\\[5pt]
We begin our analysis in the probe limit. At $\terms{\xi^0}$, we find that the terms presented in table \eqref{eq:probeAtxi0} can be generated by the following sequences 
\begin{equation}\label{eq:evenEikonalXi0}
\begin{aligned}
\delta_{\rm H,0SF}^{(2n,0)} &=
\frac{G_N^3 M^3 m\,(2 n+1) |a_2|^{2 n}}{3\, |b|^{2 (n+1)} \left(y^2-1\right)^{5/2}} \left(64 (n+1) y^6-24 (4 n+5) y^4+ (36 n+60) y^2-2 n-5\right), \\
\delta_{\rm H,0SF}^{(2n+1,0)} &=
-\frac{4G_N^3 M^3 m\, (n+1) y |a_2|^{2 n+1}}{3 |b|^{2 n+3} \left(y^2-1\right)^2}   \left(16 (2 n+3) y^4-4 (8 n+15) y^2+6 n+15\right).
\end{aligned}
\end{equation}
Summing all even and odd contributions yields the spin-resummed eikonal-like phase at order $\terms{\xi^{(0)}}$,
\begin{align}
\delta_{\rm H,0SF}^{(a_2,0)}
&= \sum_{n=0}^{\infty} \left[\delta_{\rm H}^{(2n,0)} + \delta_{\rm H}^{(2n+1,0)}\right]={G_N^3  M^3 m\over |b|^2}\nn \\
\times &\Big[-\frac{2   \left(32 y^6-48 y^4+18 y^2-1\right) \left(\beta^4+3\beta^2 \right)}{3 \left(y^2-1\right)^{5/2} \left(\beta^2-1\right)^3}+\frac{\left(64 y^6-120 y^4+60 y^2-5\right) \left(\beta^2+1\right)}{3 \left(y^2-1\right)^{5/2} \left(\beta^2-1\right)^2} \nn\\
&+{16 \beta^3  y (3 - 16 y^2 + 16 y^4)\over 3 (\beta^2 - 1)^3 (y^2-1)^2}-\frac{4 \beta   y \left(16 y^4-20 y^2+5\right)}{\left(y^2-1\right)^2 \left(\beta^2-1\right)^2}\Big]\, .
\end{align} 
We work in a similar manner for the linear and quadratic spin of the black hole.\\[5pt]
The resummed eikonal-like phase at $\terms{\xi^1}$ will then be:
\begin{align}
\delta_{\rm H,0SF}^{(a_2,1)} &=  {G^3_N M^3 m \over |b|^2}\Big[\frac{8   \left(32 y^6-48 y^4+18 y^2-1\right) \left(\beta^6+5\beta^4 \right)}{ 3 \left(y^2-1\right)^{5/2}\left(\beta^2-1\right)^4}\nn  \\
	&~~~~~~~~~~~~~~~~~~~~~~~-\frac{2  \left(96 y^6-160 y^4+70 y^2-5\right) \left(\beta^4+3\beta^2  \right)}{3 \left(y^2-1\right)^{5/2}\left(\beta^2-1\right)^3}\nn\\
	&~~-\frac{16    y \left(16 y^4-16 y^2+3\right) \left(\beta^5+\beta^3\right)}{\left(y^2-1\right)^2 \left(\beta^2-1\right)^4}+\frac{8    y \left(16 y^4-20 y^2+5\right) \left(3 \beta^3+\beta\right)}{3 \left(y^2-1\right)^2 \left(\beta^2-1\right)^3}\Big],
\end{align}
and at order $\terms{\xi^{2}}$ we find:
\begin{align}
	\delta_{\rm H,0SF}^{(a_2,2)} 
	&={G_N^3  M^3 m\over |b|^2} \Big(\frac{-4  \left(32 y^6-48 y^4+18 y^2-1\right) \left(\beta^8+10 \beta^6 +5\beta^4 \right)}{\left(y^2-1\right)^{5/2} \left(\beta^2-1\right)^5}\nn \\
	&+\frac{ \left(64 y^6-104 y^4+44 y^2-3\right) \left(\beta^6+6 \beta^4 +\beta^2\right)}{\left(y^2-1\right)^{5/2} \left(\beta^2-1\right)^4}\nn\\
	&+\frac{16   y \left(16 y^4-16 y^2+3\right) \left(3 \beta^7+5 \beta^5\right)}{\left(y^2-1\right)^2 \left(\beta^2-1\right)^5}-\frac{8    y \left(48 y^4-52 y^2+11\right) \left(\beta^5+\beta^3\right)}{\left(y^2-1\right)^2 \left(\beta^2-1\right)^4}\Big).
\end{align}
The spin-resummed scattering angle obtained from the eikonal-like phase agrees with that derived from Kerr metric scattering in \cite{Damgaard:2022jem}. It also exhibits a singularity as $\beta \to \pm 1$, as expected from the Kerr ring singularity at $|a| = |b|$.\\[5pt]
We may also consider the high-relativistic-velocity scattering regime. In this limit, the singular behavior of the eikonal-like phase as $\beta \to 1$ cancels in a nontrivial way up to order $1/y^{7}$. To this order, the eikonal exhibits singular behavior only as $\beta \to -1$:
\begin{align}
(\delta_{\rm H,0SF}^{(a_2,0)}&+\xi\delta_{\rm H, 0SF}^{(a_2,1)}+\xi^2\delta_{\rm H,0SF}^{(a_2,2)})|_{y \to \infty}={G_N^3  M^3 m\over |b|^2}\nn\\
	\times & \Big[\Big(-\frac{20 (3 \beta -2)}{3 (\beta +1)^3 y^7}-\frac{8 (6 \beta -5)}{3 (\beta +1)^3 y^5}-\frac{4 (9 \beta -10)}{3 (\beta +1)^3 y^3}+\frac{64 y}{3 (\beta +1)^3}-\frac{8 (3 \beta -5)}{3 (\beta +1)^3 y}\Big)\nn\\
	&+\xi \beta\Big(\frac{20 (3 \beta -2 )  }{(\beta +1)^4 y^7}+\frac{8 (19 \beta - 14 )  }{3 (\beta +1)^4 y^5}+\frac{4 (31 \beta -26)  }{3 (\beta +1)^4 y^3}+\frac{64 (\beta -2)  y}{3 (\beta +1)^4}+\frac{32 (\beta -1)  }{(\beta +1)^4 y}\Big)\nn\\
	&+\xi^2 \beta ^2\Big(\frac{40 (2-3 \beta ) }{(\beta +1)^5 y^7}+\frac{8 (9-13 \beta ) }{(\beta +1)^5 y^5}+\frac{8 (8-11 \beta ) }{(\beta +1)^5 y^3}+\frac{8 (7-9 \beta ) }{(\beta +1)^5 y}-\frac{64 (\beta -1)  y}{(\beta +1)^5}\Big)\Big].
\end{align}
It would be interesting to investigate whether this result continues to hold for higher spin orders of the light black hole. \\[5pt]
Beyond the probe limit, the effects on the bending angle are not yet fully understood at all spin orders, owing to the complicated structure of the entire functions appearing in the contact terms of the four-point Compton amplitude. Nevertheless, the contribution arising from the master integral $\mathcal{I}'_5$ appears to be remarkably simple, as can be seen from eqs.~\eqref{eq:c7} and \eqref{eq:c5c24} in Section~\ref{sec:1sf}.\\[5pt]
Moreover, the fact that the radiation-reaction part receives contributions only from the master integrals $\mathcal{I}'_7$ and $\mathcal{I}'_2-\mathcal{I}'_4$, suggests that this sector admits a closed-form expression that includes effects from all orders in spin. To explore this possibility, we begin by focusing on the coefficient of the master integral $\mathcal{I}'_5$, which depends only on the double-copy part of the four-point Compton amplitude. This contribution is known to all orders in spin.\\[5pt]
We calculate the coefficient $C_5$ up to $\terms{a_2^9,\xi^0}$, and we find that it contains simple structures at even and odd orders in spin:
\begin{align}
	\text{even}:&\sum_{n=0}^\infty\frac{(a_2\cdot q)^{2n}}{(2n)!},~~~~~\sum_{n=0}^\infty\frac{(\alpha)^{2n}}{(2n)!}. \nn\\
	\text{odd}:&\sum_{n=0}^\infty \frac{\left(\alpha\right)^{2n-2}}{(2 n-1)!}\, ,
\end{align}   
where $\alpha \equiv \sqrt{\alpha^2}$ and $\alpha^2$ is defined in eq.~\eqref{eq:doubleTriAmp}. These power series evaluate to the entire functions, $\cosh,\, \sinh$, and  $G_1$, that appear in the three- and four-point Compton amplitudes. The closed form expression for the coefficient of $\mathcal{I}'_5$ reads:
 \begin{align}\label{eq:C50a2}
 	C^{(a_2,0)}_5 &={64G_N^3 M^3 m^3\over \pi}  \Big[\frac{  \cosh (a_2\cdot q)}{512 \sqrt{y^2-1} \epsilon ^4}+\frac{  \left(8 y^4-8 y^2+1\right) \cosh (\alpha )}{512 \sqrt{y^2-1} \epsilon ^4}\nn\\
 	&~~~~~~~~~~~~~~~~~~~~-\frac{i y \left(2 y^2-1\right) (q\cdot S_2\cdot v_1)}{128 M  \sqrt{y^2-1} \epsilon ^4}{\sinh (\alpha ) \over \alpha} \Big]\, .
 \end{align}\\[5pt]
Importantly, the relation \eqref{eq:c5c24}, which connects $C_5$ with $C_2 - C_4$ and $C_7$, has been verified up to $\terms{a_2^5}$. Motivated by the observed resummation structure of $C_5$ in eq. \eqref{eq:C50a2}, we conjecture that our novel relation, in eq. \eqref{eq:c5c24}, holds at arbitrary order in the heavy black hole spin. Assuming this, the amplitude that contributes to the radiation-reaction part may be obtained, at all orders in $a_2$, using:
 \begin{align}
 	\mathcal{M}^{3\text{PM}}_{1\text{SF},r.r.}&= {1\over 2}(C_2-C_4) (\mathcal{I}'_{2}-\mathcal{I}'_{4})+C_7 \mathcal{I}'_{7}.
 \end{align}
The radiation-reaction contribution to the eikonal-like phase is obtained using eq.~\eqref{eq:IPSFourier} and is given by:
\begin{align}\label{eq:1sfDelta}
 \delta_{\rm H, 1SF,r.r.}^{(a_2,0)}&={64G_N^3  M^2 m^2  \over |b|^2}\Big[\frac{ y \left(2 y^2-3\right) \left(4 \left(\beta ^2+1\right) y^4-4 \left(\beta ^2+1\right) y^2+1\right) \arccosh(y)}{8  \left(\beta ^2-1\right)^2 \left(y^2-1\right)^{5/2}}\nn\\
 &~~~~~~~~~~~~~~~~~ -\frac{\left(5 y^2-8\right) \left(4 \left(\beta ^2+1\right) y^4-4 \left(\beta ^2+1\right) y^2+1\right)}{24  \left(\beta ^2-1\right)^2 \left(y^2-1\right)^2}\\
 &~~~~~~~~~~~~~~~~~+\frac{  \beta   y^2 \left(2 y^2-3\right) \left(2 y^2-1\right) \arccosh(y)}{2  \left(\beta ^2-1\right)^2 \left(y^2-1\right)^{2}}-\frac{  \beta   y \left(2 y^2-1\right) \left(5 y^2-8\right)}{6  \left(\beta ^2-1\right)^2 \left(y^2-1\right)^{3/2}}\Big]\, . \nn
 \end{align}
We may generalize the results for the coefficient of the master integral $\mathcal{I}'_{5}$ to the case where both black holes are spinning. The result may be obtained by replacing $a_2\to a_2+a_1=(1+\xi)a_2$ in eq.~\eqref{eq:C50a2}:
\begin{align}\label{eq:C5a1a2}
 	C^{(a_2,a_1)}_5 &={64G_N^3  M^3 m^3  \over \pi}\Big[\frac{  \cosh \Big((1+\xi)a_2\cdot q\Big)}{512 \sqrt{y^2-1} \epsilon ^4}+\frac{  \left(8 y^4-8 y^2+1\right) \cosh \Big((1+\xi)\alpha \Big)}{512 \sqrt{y^2-1} \epsilon ^4}\nn\\
 	&~~~~~~~~-\frac{i   y \left(2 y^2-1\right) (q\cdot S_2\cdot v_1)}{128 M  \sqrt{y^2-1} \epsilon ^4}{\sinh \Big((1+\xi)\alpha \Big) \over (1+\xi) \alpha}\Big] \, .
 \end{align} 
Combining this with the conjectured all-order validity of eq. \eqref{eq:c5c24}, we propose that the radiation-reaction contribution can be generalized to all orders in the aligned light black hole spin as follows:
 \begin{align}
 	\delta_{\rm H,1SF,r.r.}^{(a_2,a_1)}=\delta^{(a_2,0)}_{\rm H,1SF,r.r.}|_{\beta\to (1+\xi)\beta}\, . 
 \end{align}
One can also calculate the contribution to the eikonal-like phase from \( C^{(a_2,a_1)}_5 \mathcal{I}'_5 \) directly, in a similar manner to eq.~\eqref{eq:C50a2}, at all orders in spin. \\[5pt]
In the high energy limit, $y\to \infty$, we observe that the leading contribution to the conservative part comes from \( C^{(a_2,a_1)}_5 \mathcal{I}'_5 \), while the leading contribution to the radiation-reaction part comes from \( C^{(a_2,a_1)}_7 \mathcal{I}'_7 \). Using the novel relation~\eqref{eq:c5c24}, we find
\begin{align}
C_5^{(a_2,a_1)}\mathcal{I}'_5+C_7^{(a_2,a_1)}\mathcal{I}'_7&\xrightarrow[]{y\to \infty}  C_7^{(a_2,a_1)}(2\mathcal{I}'_5+\mathcal{I}'_7)\nn\\
&=C_7^{(a_2,a_1)}\Big(-2\epsilon ^3\, \arccosh(y) \mathcal{I}_{3,2}'
	+4 \epsilon ^3\, \arccosh(y) \mathcal{I}_{2,2}'+\mathcal{O}(\epsilon ^4,i) \Big)\nn\\
    &=0+\mathcal{O}(\epsilon ^4,i)\, ,
\end{align}
where we make use of the boundary-values of the master integrals in eq.~\eqref{eq:bdvalue}. We note that $C_7^{(a_2,a_1)}$ factors out in the high-energy limit and contains the information at all orders in spin. Given the assumptions discussed above, this implies that in the high energy limit, $y\to \infty$, the leading contributions from $\mathcal{I}'_5$ to the conservative dynamics should cancel against the radiation-reaction contributions from $\mathcal{I}'_7$, to all orders in spin. \\[5pt]
The observed simplifications in the radiation-reaction sector may be related to the fact that dissipative effects are expected to be strongly constrained by lower-order conservative dynamics through linear-response relations. See e.g. refs.~\cite{Bini:2012ji, Haddad:2025cmw, Alessio:2022kwv}.\\[5pt]
\section{Conclusions and Outlook}\label{sec:conclusion}
This work presents the heavy-mass effective field theory amplitude and eikonal-like phase for the gravitational scattering of two spinning black holes, incorporating contributions up to first self-force order at third post-Minkowskian order. We provide resummed results in the probe limit and in the radiation-reaction sector. Both the spin-expanded expressions and their resummed counterparts are computed and analysed in light of theoretical expectations associated with the Kerr singularity.\\[5pt]
In the zeroth self-force case, higher-spin five-point tree amplitudes are required to accurately describe the dynamics of a light black hole at even higher spin orders. The factorization properties of the massive cut enable the construction of the five-point amplitude up to fifth order in spin, provided that the corresponding Compton amplitude is available. The primary challenge arises at spin sixth order, where genuine five-point contact terms begin to contribute. These terms remain unknown, and no systematic method for constructing them for the Kerr black hole at the point of writing exists. An interesting possibility, beyond the scope of this work, is to use the Kerr action found in the hypersurface model \cite{Bjerrum-Bohr:2025lpw}, to derive such contact contributions. With these results established, a natural direction for future work would be to examine the behavior of the eikonal-like phase in the high-energy limit, particularly focusing on the potential singular structure that emerges as $\beta \to 1$ when higher-order contributions in the light black hole spin are taken into account.\\[5pt]
At first self-force order, additional contact terms in the Compton amplitude need to be introduced to fully match solutions of the Teukolsky equation \cite{Teukolsky:1973ha} in black hole perturbation theory, including the Mano-Suzuki-Takasugi solutions \cite{Mano:1996vt,Sasaki:2003xr} (See \cite{Bautista:2023szu,Bautista:2021wfy} for recent progress in tree-level matching.). Beginning at fifth order in spin, the point-particle limit of the Compton amplitude becomes incomplete, indicating that additional contact terms beyond the point-particle approximation must be included in the effective action. The corresponding contact terms will be presented in upcoming work. Consequently, the conservative contribution to the bending angle at fifth order in the background black hole spin remains partial.\\[5pt]
Extending the analysis to higher post-Minkowskian orders, particularly the fourth and fifth orders, is another avenue. Application of the heavy-mass effective field theory framework at these orders, even for low-spin or non-spinning cases, remains an open problem that needs further attention. It is of particular interest to determine whether radiation-reaction contributions or their sub-sectors continue to display a simple structure and remain free of contact terms in the generalized Compton amplitude. The essential step is to identify the spin-independent relation between the radiation-reaction (sub)-sector and the conservative part derived from the relevant master integrals. 
\begin{align}\label{masterintegraldiagram}
 \begin{tikzpicture}[baseline={([yshift=-0.2ex]current bounding box.center)}]\tikzstyle{every node}=[font=\small]	
\begin{feynman}
    	 \vertex (p1) {\(v_1,a_1\)};
    	 \vertex [right=1.5cm of p1] (b1) [dot]{};
    	  \vertex [right=3.0cm of b1] (b2) [dot]{};
    	 \vertex [right=1.2cm of b2] (p4){};
          \vertex [right=1.5cm of b1] (cutb1) []{$\textcolor{red}{\rule{1.5pt}{3ex}}$};
    	 \vertex [above=2.0cm of p1](p2){$v_2,a_2$};
    	 \vertex [right=1.5cm of p2] (u1) [dot]{};
    	 \vertex [right=1.5cm of u1] (u2) [dot]{};
         \vertex [right=1.5cm of u2] (u3) [dot]{};
    	  \vertex [right=1.2cm of u3](p3){};
           \vertex [right=0.75cm of u1] (cutu1) []{$\textcolor{red}{\rule{1.5pt}{3ex}}$};
            \vertex [right=0.75cm of u2] (cutu2) []{$\textcolor{red}{\rule{1.5pt}{3ex}}$};
    	  \vertex [above=1.cm of b1] (m1)[dot]{};
    	  \vertex [right=1.5cm of m1] (m2) [dot]{};
          \vertex [right=1.5cm of m2] (m3) [dot]{};
    	  \diagram* {(p1) -- [thick] (b1)-- [thick] (b2) -- [thick] (p4),
    	  (u1)--[thick](m1), (u3)--[thick](m3),(b1)--[thick](m1), (m2)-- [thick](u2) ,(m3)-- [thick](b2), (p2) -- [thick] (u1)-- [thick] (u2)-- [thick] (u3)-- [thick] (p3), (m1)--[thick] (m2)--[thick] (m3)
    	  };
    \end{feynman}  
    \end{tikzpicture},&&
    \begin{tikzpicture}[baseline={([yshift=-0.2ex]current bounding box.center)}]\tikzstyle{every node}=[font=\small]	
\begin{feynman}
    	 \vertex (p1) {\(v_1,a_1\)};
    	 \vertex [right=1.5cm of p1] (b1) [dot]{};
    	  \vertex [right=4.5cm of b1] (b2) [dot]{};
    	 \vertex [right=1.2cm of b2] (p4){};
          \vertex [right=2.25cm of b1] (cutb1) []{$\textcolor{red}{\rule{1.5pt}{3ex}}$};
    	 \vertex [above=2.0cm of p1](p2){$v_2,a_2$};
    	 \vertex [right=1.5cm of p2] (u1) [dot]{};
    	 \vertex [right=1.5cm of u1] (u2) [dot]{};
         \vertex [right=1.5cm of u2] (u3) [dot]{};
         \vertex [right=1.5cm of u3] (u4) [dot]{};
    	  \vertex [right=1.2cm of u4](p3){};
          \vertex [right=0.75cm of u1] (cutu1) []{$\textcolor{red}{\rule{1.5pt}{3ex}}$};
          \vertex [right=0.75cm of u2] (cutu2) []{$\textcolor{red}{\rule{1.5pt}{3ex}}$};
           \vertex [right=0.75cm of u3] (cutu3) []{$\textcolor{red}{\rule{1.5pt}{3ex}}$};
    	  \vertex [above=1.cm of b1] (m1)[dot]{};
    	  \vertex [right=1.5cm of m1] (m2) [dot]{};
          \vertex [right=1.5cm of m2] (m3) [dot]{};
          \vertex [right=1.5cm of m3] (m4) [dot]{};
    	  \diagram* {(p1) -- [thick] (b1)-- [thick] (b2) -- [thick] (p4),
    	  (u1)--[thick](m1), (u3)--[thick](m3),(u4)--[thick](m4),(b1)--[thick](m1), (m2)-- [thick](u2) ,(m4)-- [thick](b2), (p2) -- [thick] (u1)-- [thick] (u2)-- [thick] (u3)-- [thick] (u4)-- [thick] (p3), (m1)--[thick] (m2)--[thick] (m3)--[thick] (m4)
    	  };
    \end{feynman}  
    \end{tikzpicture} \,
    \end{align}
If such novel spin-independent relations exist, one may hope to derive closed-form expressions for the eikonal-like phase at all orders in spin, together with a clear understanding of their singularity structures at finite spin. We leave these and other interesting questions to future work.
\acknowledgments{
We thank P. Di Vecchia, C. J. Eriksen A. Luna, and P. L. Ortega for many insightful discussions. The work of N.E.J.B.-B., G.C. was supported in part by DFF grant 1026-00077B, The Center of Gravity, which is a Center of Excellence funded by the Danish National Research Foundation under grant No. 184. G.C. has also received funding from the European Union Horizon 2020 research and innovation program under the Marie Sklodowska-Curie grant agreement No. 847523, INTERACTIONS. }
\appendix
\section{Tensor Decomposition}\label{ap:epsdecomp}
\subsection{The General Method}
As discussed in the main text, the free Lorentz indices appearing in our tensor loop integrals are always contracted with rank-one or rank-two tensors, forming structures that can be written schematically as:
\begin{equation}
        \ell_i\cdot S_{X}\cdot p_x,  
\end{equation}
where $p_x$ can be any vector $p_x^\mu \in \{q^\mu,\, v_1^\mu,\, v_2^\mu,\, \ell_{j}^\mu\}$, and $S_X$ can be any rank-two tensor that can be constructed from our vectors and spin tensors.\\[5pt]
As discussed in Sections \ref{sec:0sf} and \ref{sec:1sf}, we resolve these structures in 4-dimensional space by decomposing $S_X$ into a basis built from the external vectors:
\begin{equation*}
v_1^\mu,\quad v_2^\mu,\quad q^\mu,\quad a_2^\mu,
\end{equation*}
In this appendix, we derive such decompositions explicitly. \\[5pt]
The method we present here is general and works for any rank-$R$ tensor, decomposed in a basis of $m$ vectors $\{X_1^\mu,\,X_2^\mu,\,\ldots\,X_m^\mu\}$ (See \cite{vanNeerven:1983vr,Anastasiou:2023koq,Ellis:2011cr} for similar decompositions.).\\[5pt]
The starting point is an ansatz containing all possible rank-$R$ tensors that can be constructed from our basis of $m$ vectors:
\begin{equation}
    S^{\mu_{1}\mu_{2}\ldots\mu_{R}}_{X} = \sum_{i_{1},i_{2},\ldots,i_{R}=1}^{m}c_{i_{1} i_{2}\ldots i_{R}}X_{i_{1}}^{\mu_{1}}X_{i_{2}}^{\mu_{2}}\ldots X_{i_{R}}^{\mu_{R}}.
\end{equation}
The coefficients $c_{i_{1} i_{2}\ldots i_{R}}$ are obtained by contracting both sides with the dual basis $\left<X_i\right>^\mu$, which is defined in terms of $G^{-1}$, the inverse of the Gram matrix $G_{kl} = X_k \cdot X_l$:
\begin{equation}
    \left<X_i\right>^\mu = \sum_{j=1}^{m}\left(G^{-1}\right)_{ij}X^{\mu}_{j}.
\end{equation}
The solution for the coefficients $c_{i_{1} i_{2}\ldots i_{R}}$ is:
\begin{equation}\label{eq:genCoefs}
    c_{i_1 i_2 \ldots i_R}
=
\langle X_{i_1}\rangle_{\mu_1}
\langle X_{i_2}\rangle_{\mu_2}
\cdots
\langle X_{i_R}\rangle_{\mu_R}\,
    S^{\mu_1\mu_2\ldots\mu_R}_{X},
\end{equation}
and the full decomposition becomes:
\begin{equation}\label{eq:tensDecomp}
S^{\mu_1,\mu_2\ldots \mu_R}_{X} = (S_{X})_{\alpha_1\alpha_2\ldots \alpha_R}\sum_{i_1,i_2,\ldots i_R=1}^{m} \langle X_{i_1}\rangle^{\alpha_1}
\langle X_{i_2}\rangle^{\alpha_2} \cdots \langle X_{i_R}\rangle^{\alpha_R}
 X_{i_1}^{\mu_1}X_{i_2}^{\mu_2}\ldots X_{i_R}^{\mu_R}.
\end{equation}
\subsection{Example: Decomposition of the Levi-Civita Tensor}
To illustrate the general method described above, we now apply it to the decomposition of the Levi–Civita tensor that appeared in the probe-limit calculation. As discussed in Section~\ref{sec:0sf}, it proved more practical to decompose the object
\begin{equation}
   S^{\mu\nu\rho} = (a_2)_{\alpha}\epsilon^{\alpha\mu\nu\rho}=\epsilon(a_2,\mu,\nu,\rho).
\end{equation}
rather than the spin tensors $S_1^{\mu\nu}$ and $S_2^{\mu\nu}$ individually.\\[5pt]
Since the tensor is totally antisymmetric in its free indices, its decomposition is restricted to antisymmetric combinations of the external vectors. We therefore write:
\begin{equation}\label{eq:epsAnsatz}
\epsilon(a_2,\mu,\nu,\rho)
    = c_1\, v_1^{[\mu}v_2^{\nu}a_2^{\rho]}
    + c_2\, v_1^{[\mu}q^{\nu}a_2^{\rho]}
    + c_3\, v_2^{[\mu}q^{\nu}a_2^{\rho]}
    + c_4\, v_1^{[\mu}v_2^{\nu}q^{\rho]}.
\end{equation}\\[5pt]
For the basis of our four external vectors $X^{\mu}_{i} = \{v_1^\mu,v_2^\mu,q^\mu,a_2^\mu\}$, the dual basis $\left<X_{i}\right>^{\mu}$, reads:
\begin{equation}
\left<X_{i}\right>^{\mu} = 
    \begin{aligned}
\Big\{ &\frac{y v_2{}^{\mu }-v_1{}^{\mu }}{y^2-1}, &
 &\frac{y v_1{}^{\mu }-v_2{}^{\mu }}{y^2-1}, &
& \frac{(a_2\cdot q) a_2{}^{\mu } - a_2^2 q^{\mu }}{\left(a_2\cdot q\right){}^2-a_2^2 q^2}, &
 &\frac{(a_2\cdot q) q^{\mu } - q^2 a_2{}^{\mu }}{\left(a_2\cdot q\right){}^2-a_2^2 q^2}\Big\}, &
    \end{aligned}
\end{equation}
and using eq.~\eqref{eq:tensDecomp}, the decomposition becomes:
\begin{equation}
\begin{aligned}
\epsilon(a_2,\mu,\nu,\rho)
    &= \frac{q\cdot S_{2}\cdot v_1 \left(\left(q\cdot a_2\right)
   v_{1}^{[\mu} v_{2}^{\nu}a_2^{\rho]} - a_2^2\, v_{1}^{[\mu} v_{2}^{\nu}q^{\rho]}\right)}{M\left(y^2-1\right) \left(\left(q\cdot a_2\right){}^2-a_2^2\,
   q^2\right)},
\end{aligned}
\end{equation}
where $q\cdot S_{2}\cdot v_1=-M\eps\left(a_2,q,v_1,v_2\right)$.\\[5pt]
After integrating over $q$, the remaining scalar product with the spin tensor is $b\cdot S_2\cdot v_1=-M\epsilon(a_2,b,v_1,v_2)$. This remaining scalar product can be simplified further by choosing an explicit coordinate system.\\[5pt]
Following~\cite{Brandhuber:2023hhl}, we work in the rest frame of the massive black hole:
\begin{equation}
\begin{aligned}
& v_2^\mu = (1,0,0,0), 
\qquad
v_1^\mu = \left( y,\ \sqrt{y^2-1},\ 0,\ 0\right),\\[4pt]
& a_2^\mu = |a_2|\,(0,0,1,0),
\qquad
b^\mu    = |b|\,(0,0,0,1),
\end{aligned}
\end{equation}
where $|a_2|>0$ and $|b|>0$ denote the magnitudes of $a_2^\mu$ and $b^\mu$, respectively. It is straightforward to check that $a_2\cdot v_2=0$ and $b\cdot v_2=0$. A direct evaluation of the Levi--Civita tensor then gives
\begin{equation}\label{eq:LCres}
\epsilon(a_2,b,v_1,v_2)
    = |a_2|\,|b|\,\sqrt{y^{2}-1}.
\end{equation}
\section{Amplitudes at first self-force order}\label{app:amp1sf}
In this appendix, we present the amplitudes in first self-force order to $\beta^5$ order, namely $A^{(j,i)}_{\rm 1SF,\bullet}, j\in [0,5-i], i\in [0,2]$, where $j$ and $i$ denote the power of $a_2$ and $\xi$. The subscript label $\bullet$ can be either ``con.'', denoting the conservative contribution, or ``r.r.'', denoting the radiation-reaction contribution. We use $p_1=m v_1$ and $a_2^2=a_2\cdot a_2, q^2=q\cdot q$ and we ignore the global coefficient $ {8\pi G_N^3 M^3 m^3\over \epsilon}({-q^2\over 2})^{-2\epsilon}$. For the radiation-reaction part,  the amplitudes have the form:
\begin{flalign}
    \begin{split}
        &A_{\rm 1SF,r.r.}^{(0,0)}=\frac{y \left(1-2 y^2\right)^2 \left(2 y^2-3\right) \, \arccosh(y)}{2 \left(y^2-1\right)^2}-\frac{\left(1-2 y^2\right)^2 \left(5 y^2-8\right)}{6 \left(y^2-1\right)^{3/2}},
        \end{split}&
\end{flalign}
\begin{flalign}
    \begin{split}
        &A^{(1,0)}_{\rm 1SF,r.r.}=\frac{-i  y \left(10 y^4-21 y^2+8\right) (v_1\cdot S_2\cdot q)}{3 M\left(y^2-1\right)^{3/2}}+\frac{i  y^2 \left(4 y^4-8 y^2+3\right) \, \arccosh(y) (v_1\cdot S_2\cdot q)}{M\left(y^2-1\right)^2},
        \end{split}&
\end{flalign}
\begin{flalign}
    \begin{split}
        &A^{(2,0)}_{\rm 1SF,r.r.}=\frac{ \left(5 y^2-8\right) \left(a_2^2 q^2 \left(8 y^4-8 y^2+1\right)-2 \left(1-2 y^2\right)^2 \left(a_2\cdot q\right){}^2\right)}{24 \left(y^2-1\right)^{3/2}}\\&+\frac{ y \left(2 y^2-3\right) \, \arccosh(y) \left(2 \left(1-2 y^2\right)^2 \left(a_2\cdot q\right){}^2+a_2^2 q^2 \left(-8 y^4+8 y^2-1\right)\right)}{8 \left(y^2-1\right)^2},
    \end{split}&
\end{flalign}
\begin{flalign}
    \begin{split}
     &A^{(3,0)}_{\rm 1SF,r.r.}=\frac{-i y^2 \left(4 y^4-8 y^2+3\right)\, \arccosh(y) \left(a_2^2 q^2-\left(q\cdot a_2\right){}^2\right) (v_1\cdot S_2\cdot q)}{6M \left(y^2-1\right)^2}\\&+\frac{i y \left(10 y^4-21 y^2+8\right) \left(a_2^2 q^2-\left(q\cdot a_2\right){}^2\right) (v_1\cdot S_2\cdot q)}{18 M\left(y^2-1\right)^{3/2}},
     \end{split}&
\end{flalign}
\begin{flalign}
    \begin{split}
        &A^{(4,0)}_{\rm 1SF,r.r.}=\frac{y \left(2 y^2-3\right) \arccosh(y)}{96 \left(y^2-1\right)^2}\bigg(a_2^4 q^4 \left(8 y^4\!-\!8 y^2\!+\!1\right)\!-\!2 a_2^2 q^2 \left(8 y^4\!-\!8 y^2\!+\!1\right) \left(q\cdot a_2\right){}^2\\
        &+2 \left(1-2 y^2\right)^2 \left(q\cdot a_2\right){}^4\bigg)-\frac{\left(5 y^2-8\right) }{288 \left(y^2-1\right)^{3/2}}\bigg(a_2^4 q^4 \left(8 y^4-8 y^2+1\right)\\&-2 a_2^2 q^2 \left(8 y^4-8 y^2+1\right) \left(q\cdot a_2\right){}^2+2 \left(1-2 y^2\right)^2 \left(q\cdot a_2\right){}^4\bigg),
        \end{split}&
\end{flalign}
\begin{flalign}
\begin{split}
    &A^{(5,0)}_{\rm 1SF,r.r.}=\frac{-i y \left(10 y^4-21 y^2+8\right) \left(\left(q\cdot a_2\right){}^2-a_2^2 q^2\right){}^2 (v_1\cdot S_2\cdot q)}{360M \left(y^2-1\right)^{3/2}}\\&+\frac{i y^2 \left(4 y^4-8 y^2+3\right) \arccosh(y) \left(\left(q\cdot a_2\right){}^2-a_2^2 q^2\right){}^2 (v_1\cdot S_2\cdot q)}{120 M\left(y^2-1\right)^2},
\end{split}&
\end{flalign}
\begin{flalign}
    \begin{split}
         &A^{(0,1)}_{\rm 1SF,r.r.}=\frac{-i  y \left(10 y^4-21 y^2+8\right) (v_1\cdot S_2\cdot q)}{3M \left(y^2-1\right)^{3/2}}+\frac{iy^2 \left(4 y^4-8 y^2+3\right) \arccosh(y) (v_1\cdot S_2\cdot q)}{M\left(y^2-1\right)^2},
          \end{split}&
\end{flalign}
\begin{flalign}
    \begin{split}
         &A^{(1,1)}_{\rm 1SF,r.r.}=\frac{ \left(5 y^2-8\right) \left(a_2^2 q^2 \left(8 y^4-8 y^2+1\right)-2 \left(1-2 y^2\right)^2 \left(a_2\cdot q\right){}^2\right)}{12 \left(y^2-1\right)^{3/2}}\\&+\frac{ y \left(2 y^2-3\right) \arccosh(y) \left(2 \left(1-2 y^2\right)^2 \left(a_2\cdot q\right){}^2+a_2^2 q^2 \left(-8 y^4+8 y^2-1\right)\right)}{4 \left(y^2-1\right)^2},
         \end{split}&
\end{flalign}
\begin{flalign}
    \begin{split}
       &A^{(2,1)}_{\rm 1SF,r.r.}=\frac{-i y^2 \left(4 y^4-8 y^2+3\right) \arccosh(y) \left(a_2^2 q^2-\left(q\cdot a_2\right){}^2\right) (v_1\cdot S_2\cdot q)}{2M \left(y^2-1\right)^2}\\&+\frac{i y \left(10 y^4-21 y^2+8\right) \left(a_2^2 q^2-\left(q\cdot a_2\right){}^2\right) (v_1\cdot S_2\cdot q)}{6M \left(y^2-1\right)^{3/2}},
       \end{split}&
\end{flalign}
\begin{flalign}
    \begin{split}
         &A^{(3,1)}_{\rm 1SF,r.r.}=\frac{y \left(2 y^2-3\right) \arccosh(y) }{24 \left(y^2-1\right)^2}\bigg(a_2^4 q^4 \left(8 y^4-8 y^2+1\right)-2 a_2^2 q^2 \left(8 y^4-8 y^2+1\right) \left(q\cdot a_2\right){}^2\\&+2 \left(1-2 y^2\right)^2 \left(q\cdot a_2\right){}^4\bigg)-\frac{\left(5 y^2-8\right) }{72 \left(y^2-1\right)^{3/2}}\bigg(a_2^4 q^4 \left(8 y^4-8 y^2+1\right)\\&-2 a_2^2 q^2 \left(8 y^4-8 y^2+1\right) \left(q\cdot a_2\right){}^2+2 \left(1-2 y^2\right)^2 \left(q\cdot a_2\right){}^4\bigg),
    \end{split}&
\end{flalign}
\begin{flalign}
    \begin{split}
         &A^{(4,1)}_{\rm 1SF,r.r.}=\frac{-i y \left(10 y^4-21 y^2+8\right) \left(\left(q\cdot a_2\right){}^2-a_2^2 q^2\right){}^2 (v_1\cdot S_2\cdot q)}{72 M\left(y^2-1\right)^{3/2}}\\&+\frac{i y^2 \left(4 y^4-8 y^2+3\right) \arccosh(y) \left(\left(q\cdot a_2\right){}^2-a_2^2 q^2\right){}^2 (v_1\cdot S_2\cdot q)}{24M \left(y^2-1\right)^2}, 
    \end{split}&
\end{flalign}
\begin{flalign}
    \begin{split}
         &A^{(0,2)}_{\rm 1SF,r.r.}=\frac{ \left(5 y^2-8\right) \left(a_2^2 q^2 \left(8 y^4-8 y^2+1\right)-2 \left(1-2 y^2\right)^2 \left(a_2\cdot q\right){}^2\right)}{24 \left(y^2-1\right)^{3/2}}\\&+\frac{y \left(2 y^2-3\right) \arccosh(y) \left(2 \left(1-2 y^2\right)^2 \left(a_2\cdot q\right){}^2+a_2^2 q^2 \left(-8 y^4+8 y^2-1\right)\right)}{8 \left(y^2-1\right)^2},
    \end{split}&
\end{flalign}
\begin{flalign}
     \begin{split}
      &A^{(1,2)}_{\rm 1SF,r.r.}=\frac{-i y^2 \left(4 y^4-8 y^2+3\right) \arccosh(y) \left(a_2^2 q^2-\left(q\cdot a_2\right){}^2\right) (v_1\cdot S_2\cdot q)}{2 M\left(y^2-1\right)^2}\\&+\frac{i y \left(10 y^4-21 y^2+8\right) \left(a_2^2 q^2-\left(q\cdot a_2\right){}^2\right) (v_1\cdot S_2\cdot q)}{6 M\left(y^2-1\right)^{3/2}},
      \end{split}&
\end{flalign}
\begin{flalign}
     \begin{split}
    &A^{(2,2)}_{\rm 1SF,r.r.}=\!\frac{y \left(2 y^2\!-\!3\right) \arccosh(y) }{16 \left(y^2-1\right)^2}\bigg(a_2^4 q^4 \left(8 y^4\!-\!8 y^2\!+\!1\right)-2 a_2^2 q^2 \left(8 y^4\!-\!8 y^2\!+\!1\right) \left(q\cdot a_2\right){}^2\\&+2 \left(1-2 y^2\right)^2 \left(q\cdot a_2\right){}^4\bigg)-\frac{\left(5 y^2-8\right) }{48 \left(y^2-1\right)^{3/2}}\bigg(a_2^4 q^4 \left(8 y^4-8 y^2+1\right)\\&-2 a_2^2 q^2 \left(8 y^4-8 y^2+1\right) \left(q\cdot a_2\right){}^2+2 \left(1-2 y^2\right)^2 \left(q\cdot a_2\right){}^4\bigg),
    \end{split}&
\end{flalign}
\begin{flalign}
     \begin{split}
         &A^{(3,2)}_{\rm 1SF,r.r.}=\frac{-i y \left(10 y^4-21 y^2+8\right) \left(\left(q\cdot a_2\right){}^2-a_2^2 q^2\right){}^2 (v_1\cdot S_2\cdot q)}{36M \left(y^2-1\right)^{3/2}}\\&+\frac{i y^2 \left(4 y^4-8 y^2+3\right) \arccosh(y) \left(\left(q\cdot a_2\right){}^2-a_2^2 q^2\right){}^2 (v_1\cdot S_2\cdot q)}{12M \left(y^2-1\right)^2}\, .
    \end{split}&
\end{flalign}
For the conservative part,  the amplitudes are listed below. At $\terms{\xi^0}$ order, the conservative part of the amplitudes agrees with \cite{Akpinar:2025bkt} up to $a_2^4$ for the spin-aligned configuration.
\begin{flalign}
     \begin{split}
        &A^{(0,0)}_{\rm 1SF,con.}=\frac{y \left(6 y^2 \left(6 y^4-19 y^2+22\right)-55\right)}{6 \left(y^2-1\right)^2}+\frac{\left(-4 y^4+12 y^2+3\right) \arccosh(y)}{ \sqrt{y^2-1}},
        \end{split}&
    \end{flalign}
\begin{flalign}
     \begin{split}
        &A^{(1,0)}_{\rm 1SF,con.}=\frac{-2i y \left(y^2-6\right) \left(2 y^2+1\right) \arccosh(y) (v_1\cdot S_2\cdot q)}{M\left(y^2-1\right)^{3/2}}\\&+\frac{i  \left(36 y^6-156 y^4+84 y^2+41\right) (v_1\cdot S_2\cdot q)}{6M \left(y^2-1\right)^2}, 
    \end{split}&
\end{flalign}
    \begin{flalign}
        \begin{split}
        &A^{(2,0)}_{\rm 1SF,con.}=\frac{y}{60 \left(y^2-1\right)^2} \bigg(\big(2 \left(90 y^4-621 y^2+52\right) y^2+213\big) \left(q\cdot a_2\right){}^2\\&-a_2^2 q^2 \left(186 y^6-1257 y^4+446 y^2+1005\right)\bigg) +\frac{\arccosh(y) }{4 \left(y^2-1\right)^{5/2}}\\&\times\bigg(a_2^2 q^2 \left(8 y^8-72 y^6+11 y^4+66 y^2+12\right)-2 y^2 \left(4 y^6-36 y^4+y^2+6\right) \left(q\cdot a_2\right){}^2\bigg),
        \end{split}&
\end{flalign}
\begin{flalign}
     \begin{split}
        &A^{(3,0)}_{\rm 1SF,con.}=\frac{-i  (v_1\cdot S_2\cdot q)}{180 M\left(y^2-1\right)^3}\bigg(a_2^2 q^2 \left(192 y^8-1780 y^6+251 y^4+2930 y^2+297\right)\\&+\left(-180 y^8+1744 y^6+904 y^4+985 y^2+327\right) \left(q\cdot a_2\right){}^2\bigg)\\&+\frac{i y\, \arccosh(y) ( v_1\cdot S_2\cdot q)}{12M \left(y^2-1\right)^{7/2}}\bigg(a_2^2 q^2 \left(8 y^8-92 y^6-10 y^4+141 y^2+79\right)\\&-4 \left(2 y^2+1\right) \left(y^6-12 y^4+y^2-11\right) \left(q\cdot a_2\right){}^2\bigg),
         \end{split}&
\end{flalign}
\begin{flalign}
     \begin{split}
        &A^{(4,0)}_{\rm 1SF,con.}=\frac{y}{10080 \left(y^2-1\right)^4}\bigg(a_2^4 q^4 \big(2772 y^{10}-35530 y^8+29980 y^6+93672 y^4-68384 y^2\\&-29125\big)-4 a_2^2 q^2 \left(1326 y^{10}-17495 y^8+6320 y^6+15441 y^4+278 y^2+7360\right) \left(q\cdot a_2\right){}^2\\&+2 \left(1260 y^{10}-17198 y^8-2188 y^6-17487 y^4+14150 y^2+12643\right) \left(q\cdot a_2\right){}^4\bigg)\\&+\frac{\arccosh(y)}{96 \left(y^2-1\right)^{9/2}}\bigg(a_2^4 q^4 \left(25-2 y^2 \left(8 y^{10}-120 y^8+69 y^6+300 y^4+15 y^2-291\right)\right)\\&+4 a_2^2 q^2 \left(8 y^{12}-120 y^{10}+39 y^8+36 y^6+96 y^4+54 y^2+13\right) \left(q\cdot a_2\right){}^2\\&-4 \left(4 y^{12}-60 y^{10}+5 y^8-116 y^6+30 y^4+88 y^2+7\right) \left(q\cdot a_2\right){}^4\bigg),
         \end{split}&
\end{flalign}
\begin{flalign}
     \begin{split}
        &A^{(5,0)}_{\rm 1SF,con.}=\frac{-i y \, \arccosh(y)  (v_1\cdot S_2\cdot q)}{3840M \left(y^2-1\right)^{11/2}}\bigg(a_2^4 q^4 \big(4 y^2 (32 y^{10}-560 y^8-4 y^6-30 y^4-1851 y^2\\&-3383)-1323\big)-8 a_2^2 q^2 \big(6755 y^2+4 (8 y^8-140 y^6+84 y^4+1245 y^2+2972) y^4\\&+1076\big) \left(q\cdot a_2\right){}^2+8 \big(-2166 y^2+8 \left(2 y^8-35 y^6-14 y^4-219 y^2-452\right) y^4\\&-259\big) \left(q\cdot a_2\right){}^4\bigg)+\frac{i (v_1\cdot S_2\cdot q)}{403200M \left(y^2-1\right)^5}\bigg(a_2^4 q^4 \big(-823341 y^2+2 (11848 y^8-178280 y^6\\&+21386 y^4+96430 y^2-814757) y^4-23148\big)-8 a_2^2 q^2 \big(5512 y^{12}-87024 y^{10}+102210 y^8\\&+905378 y^6+1153333 y^4+469888 y^2+23938\big) \left(q\cdot a_2\right){}^2+8 \big(2520 y^{12}-42304 y^{10}\\&-32144 y^8-247742 y^6-409434 y^4-121475 y^2-7166\big) \left(q\cdot a_2\right){}^4\bigg),
    \end{split}&
\end{flalign}
\begin{flalign}
     \begin{split}
        &A^{(0,1)}_{\rm 1SF,con.}=\frac{-2 i y \left(y^2-6\right) \left(2 y^2+1\right) \arccosh(y) (v_1\cdot S_2\cdot q)}{M\left(y^2-1\right)^{3/2}}\\&+\frac{i \left(36 y^6-156 y^4+84 y^2+41\right) (v_1\cdot S_2\cdot q)}{6 M\left(y^2-1\right)^2},
        \end{split}&
\end{flalign}
    \begin{flalign}
     \begin{split}
        &A^{(1,1)}_{\rm 1SF,con.}=\frac{\arccosh(y) }{8 \left(y^2-1\right)^{5/2}}\bigg(a_2^2 q^2 \left(4 \left(8 \left(y^6-10 y^4+y^2\right)+87\right) y^2+61\right)\\&+2 \left(1-4 y^2 \left(4 y^6-40 y^4-5 y^2+3\right)\right) \left(q\cdot a_2\right){}^2\bigg)\\&-\frac{y}{24 \left(y^2-1\right)^2}\bigg(a_2^2 q^2 \left(144 y^6-1136 y^4+426 y^2+1029\right)\\&+2 \left(-72 y^6+568 y^4+30 y^2-69\right) \left(q\cdot a_2\right){}^2\bigg),
    \end{split}&
\end{flalign}
\begin{flalign}
     \begin{split}
       &A^{(2,1)}_{\rm 1SF,con.}=\frac{-i  (v_1\cdot S_2\cdot q)}{120M \left(y^2-1\right)^3}\bigg(a_2^2 q^2 \left(372 y^8-4348 y^6+431 y^4+9054 y^2+956\right)\\&+2 \left(4 \left(-45 y^6+539 y^4+514 y^2+500\right) y^2+433\right) \left(q\cdot a_2\right){}^2\bigg)\\&+\frac{i y \, \arccosh(y)  (v_1\cdot S_2\cdot q)}{8M \left(y^2-1\right)^{7/2}}\bigg(a_2^2 q^2 \left(16 y^8-216 y^6-66 y^4+453 y^2+244\right)\\&+2 \left(-8 y^8+108 y^6+96 y^4+168 y^2+67\right) \left(q\cdot a_2\right){}^2\bigg),
        \end{split}&
\end{flalign}
\begin{flalign}
     \begin{split}
        &A^{(3,1)}_{\rm 1SF,con.}=\frac{\arccosh(y) }{768 \left(y^2-1\right)^{9/2}}\bigg(a_2^4 q^4 \big(33636 y^2\!-\!32 \left(16 y^8\!-\!288 y^6\!+\!66 y^4\!+\!1125 y^2\!-\!9\right) y^4\\&+1667\big)+8 a_2^2 q^2 \big(2421 y^2+8 \left(16 y^8-288 y^6-66 y^4+165 y^2+603\right) y^4+322\big) \left(q\cdot a_2\right){}^2\\&-8 \left(948 y^4+2326 y^2+16 \left(4 y^6-72 y^4-49 y^2-227\right) y^6+169\right) \left(q\cdot a_2\right){}^4\bigg)\\&+\frac{y}{11520 \left(y^2-1\right)^4}\bigg(a_2^4 q^4 \big(-550150 y^2+16 \left(768 y^6-12728 y^4+9167 y^2+47298\right) y^4\\&-254675\big)-8 a_2^2 q^2 \big(2976 y^{10}-50480 y^8-4000 y^6+70614 y^4+43457 y^2\\&+30178\big) \left(q\cdot a_2\right){}^2+8 \big(4 \left(360 y^8-6256 y^6-5546 y^4-10098 y^2+8263\right) y^2\\&+22193\big) \left(q\cdot a_2\right){}^4\bigg),
          \end{split}&
\end{flalign}
\begin{flalign}
     \begin{split}
        &A^{(4,1)}_{\rm 1SF,con.}=\frac{-i y \, \arccosh(y)( v_1\cdot S_2\cdot q)}{768 M\left(y^2-1\right)^{11/2}}\bigg(a_2^4 q^4 \big(16 \left(8 y^6\!-\!172 y^4\!-\!38 y^2\!+\!1005\right) y^6\!+\!6298 y^4\\&-26575 y^2-4289\big)-4 a_2^2 q^2 \big(2 \left(32 y^8-688 y^6-604 y^4-162 y^2+5071\right) y^4+12977 y^2\\&+3161\big) \left(q\cdot a_2\right){}^2+16 \big(327 y^2+2 \left(4 y^8-86 y^6-132 y^4-648 y^2-301\right) y^4\\&+46\big) \left(q\cdot a_2\right){}^4\bigg)+\frac{i  (v_1\cdot S_2\cdot q)}{80640M \left(y^2-1\right)^5}\bigg(a_2^4 q^4 \big(-1606275 y^4-2022279 y^2\\&+4 \left(5544 y^6-113496 y^4+44103 y^2+686150\right) y^6-91040\big)-4 a_2^2 q^2 \big(1250361 y^4\\&+1148999 y^2+4 \left(2652 y^6-56148 y^4-41372 y^2+91059\right) y^6+76656\big) \left(q\cdot a_2\right){}^2\\&+16 \big(1260 y^{12}-27744 y^{10}-52556 y^8-138773 y^6-20388 y^4\\&+33659 y^2-523\big) \left(q\cdot a_2\right){}^4\bigg),
    \end{split}&
\end{flalign}
\begin{flalign}
     \begin{split}
        &A^{(0,2)}_{\rm 1SF,con.}= \frac{y }{60 \left(y^2-1\right)^2}\bigg(\left(2 \left(90 y^4-621 y^2+52\right) y^2+213\right) \left(q\cdot a_2\right){}^2\\&-a_2^2 q^2 \left(186 y^6-1257 y^4+446 y^2+1005\right)\bigg)+\frac{\arccosh(y) }{4 \left(y^2-1\right)^{5/2}}\\&\times\bigg(a_2^2 q^2 \left(8 y^8-72 y^6+11 y^4+66 y^2+12\right)-2 y^2 \left(4 y^6-36 y^4+y^2+6\right) \left(q\cdot a_2\right){}^2\bigg),
    \end{split}&
\end{flalign}
\begin{flalign}
     \begin{split}
       &A^{(1,2)}_{\rm 1SF,con.}= \frac{-i  (v_1\cdot S_2\cdot q)}{120M \left(y^2-1\right)^3}\bigg(a_2^2 q^2 \left(372 y^8-4348 y^6+431 y^4+9054 y^2+956\right)\\&+2 \left(4 \left(-45 y^6+539 y^4+514 y^2+500\right) y^2+433\right) \left(q\cdot a_2\right){}^2\bigg)+\frac{i y \, \arccosh(y)  (v_1\cdot S_2\cdot q)}{8M \left(y^2-1\right)^{7/2}}\\&\times\bigg(a_2^2 q^2 \left(16 y^8-216 y^6-66 y^4+453 y^2+244\right)+2 \big(-8 y^8+108 y^6+96 y^4\\&+168 y^2+67\big) \left(q\cdot a_2\right){}^2\bigg),
      \end{split}&
\end{flalign}
\begin{flalign}
     \begin{split}
       &A^{(2,2)}_{\rm 1SF,con.}=\frac{y }{960 \left(y^2-1\right)^4}\bigg(a_2^4 q^4 \big(2 \left(768 y^6-13696 y^4+8956 y^2+56543\right) y^4-82545 y^2\\&-39472\big)-8 a_2^2 q^2 \big(372 y^{10}-6794 y^8-1718 y^6+11353 y^4+9009 y^2+4653\big) \left(q\cdot a_2\right){}^2\\&+8 \left(180 y^{10}-3370 y^8-3992 y^6-6489 y^4+4879 y^2+3167\right) \left(q\cdot a_2\right){}^4\bigg)\\&+\frac{\arccosh(y) }{64 \left(y^2-1\right)^{9/2}}\bigg(a_2^4 q^4 \left(5121 y^2-4 \left(16 y^8-304 y^6+24 y^4+1354 y^2-25\right) y^4+264\right)\\&+8 a_2^2 q^2 \left(16 y^{12}-304 y^{10}-138 y^8+220 y^6+865 y^4+419 y^2+47\right) \left(q\cdot a_2\right){}^2\\&-8 \left(337 y^2+2 \left(4 y^8-76 y^6-75 y^4-284 y^2+63\right) y^4+24\right) \left(q\cdot a_2\right){}^4\bigg),
         \end{split}&
\end{flalign}
\begin{flalign}
     \begin{split}
       &A^{(3,2)}_{\rm 1SF,con.}=\frac{-i y \, \arccosh(y)  (v_1\cdot S_2\cdot q)}{384 M\left(y^2-1\right)^{11/2}}\bigg(a_2^4 q^4 \big(16 \left(8 y^6\!-\!188 y^4\!-\!106 y^2\!+\!1329\right) y^6\!+\!7710 y^4\\&-37899 y^2-6533\big)-8 a_2^2 q^2 \big(32 y^{12}-752 y^{10}-1038 y^8-57 y^6+8937 y^4+10625 y^2\\&+2287\big) \left(q\cdot a_2\right){}^2+8 \big(16 y^{12}-376 y^{10}-832 y^8-3768 y^6-2274 y^4+473 y^2\\&+83\big) \left(q\cdot a_2\right){}^4\bigg)+\frac{i  (v_1\cdot S_2\cdot q)}{5760M \left(y^2-1\right)^5}\bigg(a_2^4 q^4 \big(3168 y^{12}-73056 y^{10}+14580 y^8+517700 y^6\\&-318697 y^4-424069 y^2-20136\big)-8 a_2^2 q^2 \big(756 y^{12}-18084 y^{10}-20809 y^8+42291 y^6\\&+162923 y^4+125691 y^2+7742\big) \left(q\cdot a_2\right){}^2+8 \big(360 y^{12}-8952 y^{10}-23032 y^8-60442 y^6\\&-16811 y^4+8921 y^2-214\big) \left(q\cdot a_2\right){}^4\bigg).
    \end{split}&
\end{flalign}

\section{The eikonal-like phase at first self-force order}\label{app:chi} 
At $\xi^{0},\xi,\xi^{2}$ orders, the eikonal-like phase up to $\terms{\beta^5}$  are:
\begin{align}\label{eq:chi0even_i0}
\delta^{(0,0)}_{\rm H, 1SF}
&=\frac{8 G_N^3 M^2 m^2}{|b|^2}\times\nn\\
\Bigg (&
\left[
\frac{y \left(6 y^2 \left(6 y^4-19 y^2+22\right)-55\right)}{3 \left(y^2-1\right)^{5/2}}
+\frac{\left(-8 y^4+24 y^2+6\right)\arccosh(y)}{\left(y^2-1\right)}
\right]_{\rm con.}\nn\\
&+
\left[
\frac{y \left(1-2 y^2\right)^2 \left(2 y^2-3\right)\arccosh(y)}{\left(y^2-1\right)^{5/2}}
-\frac{\left(1-2 y^2\right)^2 \left(5 y^2-8\right)}{3 \left(y^2-1\right)^2}
\right]_{\rm r.r.}
\Bigg ),\\
\delta^{(1,0)}_{\rm H, 1SF}
&=\frac{8 G_N^3 M^2 m^2}{|b|^2}\times\nn\\
\Bigg (&
\left[
-\frac{2 \beta \left(36 y^6-156 y^4+84 y^2+41\right)}{3 \left(y^2-1\right)^2}
+\frac{8 \beta y \left(y^2-6\right)\left(2 y^2+1\right)\arccosh(y)}{\left(y^2-1\right)^{3/2}}
\right]_{\rm con.}\nn\\
&+
\left[
-\frac{4 \beta y^2 \left(4 y^4-8 y^2+3\right)\arccosh(y)}{\left(y^2-1\right)^2}
+\frac{4 \beta y \left(10 y^4-21 y^2+8\right)}{3 \left(y^2-1\right)^{3/2}}
\right]_{\rm r.r.}
\Bigg),\\
\delta^{(2,0)}_{\rm H, 1SF}
&=\frac{8 G_N^3 M^2 m^2}{|b|^2}\times\nn\\
\Bigg(&
\left[
\frac{\beta^2 y \left(4 \left(46 y^4-313 y^2+83\right) y^2+741\right)}{5 \left(y^2-1\right)^{5/2}}
-\frac{24 \beta^2 \left(y^6-8 y^4-7 y^2-1\right)\arccosh(y)}{\left(y^2-1\right)^2}
\right]_{\rm con.}\nn\\
&+
\left[
\frac{2 \beta^2 y \left(2 y^2-3\right)\left(6 y^4-6 y^2+1\right)\arccosh(y)}{\left(y^2-1\right)^{5/2}}\right]_{\rm r.r.}\nn\\
&+\left[\frac{2 \beta^2 \left(-30 y^6+78 y^4-53 y^2+8\right)}{3 \left(y^2-1\right)^2}
\right]_{\rm r.r.}
\Bigg),\\
\delta^{(3,0)}_{\rm H, 1SF}
&=\frac{8 G_N^3 M^2 m^2}{|b|^2}\times\nn\\
\Bigg(&
\left[
-\frac{4 \beta^3 \left(2 y^2 \left(94 y^4-790 y^2-857\right)-89\right)}{15 \left(y^2-1\right)^2}\right]_{\rm con.}\nn\\
&-\left[\frac{8 \beta^3 y \left(-4 y^6+42 y^4+52 y^2+19\right)\arccosh(y)}{\left(y^2-1\right)^{5/2}}
\right]_{\rm con.}\nn\\
&+
\left[
\frac{8 \beta^3 y \left(10 y^4-21 y^2+8\right)}{3 \left(y^2-1\right)^{3/2}}
-\frac{8 \beta^3 y^2 \left(4 y^4-8 y^2+3\right)\arccosh(y)}{\left(y^2-1\right)^2}
\right]_{\rm r.r.}
\Bigg),\\
\delta^{(4,0)}_{\rm H, 1SF}
&=\frac{8 G_N^3 M^2 m^2}{|b|^2}\times\nn\\
\Bigg(&
\left[
\frac{\beta^4 y \left(6724 y^6-74450 y^4-109412 y^2-16377\right)}{105 \left(y^2-1\right)^{5/2}}\right]_{\rm con.}\nn\\
&+\left[\frac{2 \beta^4 \left(-20 y^8+260 y^6+417 y^4+266 y^2+1\right)\arccosh(y)}{\left(y^2-1\right)^3}
\right]_{\rm con.}\nn\\
&+
\left[
\frac{\beta^4 y \left(40 y^6-100 y^4+66 y^2-9\right)\arccosh(y)}{\left(y^2-1\right)^{5/2}}\right]_{\rm r.r.}\nn\\
&-\left[\frac{\beta^4 \left(5 y^2-8\right)\left(20 y^4-20 y^2+3\right)}{3 \left(y^2-1\right)^2}
\right]_{\rm r.r.}
\Bigg),\\
\delta^{(5,0)}_{\rm H, 1SF}
&=\frac{8 G_N^3 M^2 m^2}{|b|^2}\times\nn\\
\Bigg(&
\left[
-\frac{2 \beta^5 \left(4228 y^8-57296 y^6-103744 y^4-24153 y^2+323\right)}{105 \left(y^2-1\right)^3}
\right]_{\rm con.}\nn\\
&+
\left[
-\frac{4 \beta^5 y \left(-60 y^8+930 y^6+1840 y^4+1604 y^2-13\right)\arccosh(y)}{5 \left(y^2-1\right)^{7/2}}
\right]_{\rm con.}\nn\\
&+
\left[
\frac{4 \beta^5 y \left(10 y^4-21 y^2+8\right)}{\left(y^2-1\right)^{3/2}}
-\frac{12 \beta^5 y^2 \left(4 y^4-8 y^2+3\right)\arccosh(y)}{\left(y^2-1\right)^2}
\right]_{\rm r.r.}
\Bigg).
\end{align}

\begin{align}\label{eq:chi1even_i1}
\delta^{(0,1)}_{\rm H, 1SF}
&=\frac{8 G_N^3 M^2 m^2 \xi}{|b|^2}\times\nn\\
\Bigg(&
\left[
-\frac{2 \beta \left(36 y^6-156 y^4+84 y^2+41\right)}{3 \left(y^2-1\right)^2}
+\frac{8 \beta y \left(y^2-6\right)\left(2 y^2+1\right)\arccosh(y)}{\left(y^2-1\right)^{3/2}}
\right]_{\rm con.}\nn\\
&+
\left[
-\frac{4 \beta y^2 \left(4 y^4-8 y^2+3\right)\arccosh(y)}{\left(y^2-1\right)^2}
+\frac{4 \beta y \left(10 y^4-21 y^2+8\right)}{3 \left(y^2-1\right)^{3/2}}
\right]_{\rm r.r.}
\Bigg),\\
\delta^{(1,1)}_{\rm H, 1SF}
&=\frac{8 G_N^3 M^2 m^2 \xi}{|b|^2}\times\nn\\
\Bigg(&
\left[
\frac{2 \beta^2 y \left(36 y^6-284 y^4+66 y^2+183\right)}{\left(y^2-1\right)^{5/2}}\right]_{\rm con.}\nn\\
&+\left[\frac{12 \beta^2 \left(-4 y^6+36 y^4+35 y^2+5\right)\arccosh(y)}{\left(y^2-1\right)^2}
\right]_{\rm con.}\nn\\
&+
\left[
\frac{4 \beta^2 y \left(2 y^2-3\right)\left(6 y^4-6 y^2+1\right)\arccosh(y)}{\left(y^2-1\right)^{5/2}}\right]_{\rm r.r.}\nn\\
&-\left[\frac{4 \beta^2 \left(5 y^2-8\right)\left(6 y^4-6 y^2+1\right)}{3 \left(y^2-1\right)^2}
\right]_{\rm r.r.}
\Bigg),\\
\delta^{(2,1)}_{\rm H, 1SF}
&=\frac{8 G_N^3 M^2 m^2 \xi}{|b|^2}\times\nn\\
\Bigg(&
\left[
-\frac{4 \beta^3 \left(552 y^6-5952 y^4-7577 y^2-523\right)}{15 \left(y^2-1\right)^2}\right]_{\rm con.}\nn\\
&-\left[\frac{12 \beta^3 y \left(-8 y^6+100 y^4+154 y^2+59\right)\arccosh(y)}{\left(y^2-1\right)^{5/2}}
\right]_{\rm con.}\nn\\
&+
\left[
\frac{8 \beta^3 y \left(10 y^4-21 y^2+8\right)}{\left(y^2-1\right)^{3/2}}
-\frac{24 \beta^3 y^2 \left(4 y^4-8 y^2+3\right)\arccosh(y)}{\left(y^2-1\right)^2}
\right]_{\rm r.r.}
\Bigg),\\
\delta^{(3,1)}_{\rm H, 1SF}
&=\frac{8 G_N^3 M^2 m^2 \xi}{|b|^2}\times\nn\\
\Bigg(&
\left[
\frac{2 \beta^4 y \left(376 y^6-5576 y^4-10326 y^2-2129\right)}{3 \left(y^2-1\right)^{5/2}}
\right]_{\rm con.}\nn\\
&+\left[\frac{2 \beta^4 \left(4 \left(-20 y^6+320 y^4+687 y^2+476\right) y^2+43\right)\arccosh(y)}{\left(y^2-1\right)^3}
\right]_{\rm con.}\nn\\
&+
\left[
\frac{4 a_2^4 y \left(40 y^6-100 y^4+66 y^2-9\right)\arccosh(y)}{b^6 \left(y^2-1\right)^{5/2}}\right]_{\rm r.r.}\nn\\
&-\left[\frac{4 a_2^4 \left(5 y^2-8\right)\left(20 y^4-20 y^2+3\right)}{3 b^6 \left(y^2-1\right)^2}
\right]_{\rm r.r.}
\Bigg),\\
\delta^{(4,1)}_{\rm H, 1SF}
&=\frac{8 G_N^3 M^2 m^2 \xi}{|b|^2}\times\nn\\
\Bigg(&
\left[
-\frac{4 \beta^5 y \left(5 \left(-12 y^6+234 y^4+650 y^2+667\right) y^2+213\right)\arccosh(y)}{\left(y^2-1\right)^{7/2}}
\right]_{\rm con.}\nn\\
&+
\left[
-\frac{2 \beta^5 \left(20172 y^8-382176 y^6-936730 y^4-353185 y^2-8761\right)}{105 \left(y^2-1\right)^3}
\right]_{\rm con.}\nn\\
&+
\left[
\frac{20 \beta^5 y \left(10 y^4-21 y^2+8\right)}{\left(y^2-1\right)^{3/2}}
-\frac{60 \beta^5 y^2 \left(4 y^4-8 y^2+3\right)\arccosh(y)}{\left(y^2-1\right)^2}
\right]_{\rm r.r.}
\Bigg).
\end{align}

\begin{align}\label{eq:chi2even_i0}
\delta^{(0,2)}_{\rm H, 1SF}
&=\frac{8 G_N^3 M^2 m^2 \xi^2}{|b|^2}\times\nn\\
\Bigg(&
\left[
\frac{\beta^2 y \left(4 \left(46 y^4-313 y^2+83\right) y^2+741\right)}{5 \left(y^2-1\right)^{5/2}}\right]_{\rm con.}\nn\\
&-\left[\frac{24 \beta^2 \left(y^6-8 y^4-7 y^2-1\right)\arccosh(y)}{\left(y^2-1\right)^2}
\right]_{\rm con.}\nn\\
&+
\left[
\frac{2 \beta^2 y \left(2 y^2-3\right)\left(6 y^4-6 y^2+1\right)\arccosh(y)}{\left(y^2-1\right)^{5/2}}\right]_{\rm r.r.}\nn\\
&+\left[\frac{2 \beta^2 \left(-30 y^6+78 y^4-53 y^2+8\right)}{3 \left(y^2-1\right)^2}
\right]_{\rm r.r.}
\Bigg),\\
\delta^{(1,2)}_{\rm H, 1SF}
&=\frac{8 G_N^3 M^2 m^2 \xi^2}{|b|^2}\times\nn\\
\Bigg(&
\left[
-\frac{4 \beta^3 \left(552 y^6-5952 y^4-7577 y^2-523\right)}{15 \left(y^2-1\right)^2}\right]_{\rm con.}\nn\\
&-\left[\frac{12 \beta^3 y \left(-8 y^6+100 y^4+154 y^2+59\right)\arccosh(y)}{\left(y^2-1\right)^{5/2}}
\right]_{\rm con.}\nn\\
&+
\left[
\frac{8 \beta^3 y \left(10 y^4-21 y^2+8\right)}{\left(y^2-1\right)^{3/2}}
-\frac{24 \beta^3 y^2 \left(4 y^4-8 y^2+3\right)\arccosh(y)}{\left(y^2-1\right)^2}
\right]_{\rm r.r.}
\Bigg),\\
\delta^{(2,2)}_{\rm H, 1SF}
&=\frac{8 G_N^3 M^2 m^2 \xi^2}{|b|^2}\times\nn\\
\Bigg(&
\left[
\frac{2 \beta^4 y \left(564 y^6-9090 y^4-18244 y^2-4133\right)}{3 \left(y^2-1\right)^{5/2}}
\right]_{\rm con.}\nn\\
&+\left[\frac{4 \beta^4 \left(3 \left(-20 y^6+340 y^4+805 y^2+578\right) y^2+49\right)\arccosh(y)}{\left(y^2-1\right)^3}
\right]_{\rm con.}\nn\\
&+
\left[
\frac{6 \beta^4 y \left(40 y^6-100 y^4+66 y^2-9\right)\arccosh(y)}{\left(y^2-1\right)^{5/2}}\right]_{\rm r.r.}\nn\\
&-\left[\frac{2 \beta^4 \left(5 y^2-8\right)\left(20 y^4-20 y^2+3\right)}{\left(y^2-1\right)^2}
\right]_{\rm r.r.}
\Bigg),\\
\delta^{(3,2)}_{\rm H, 1SF}
&=\frac{8 G_N^3 M^2 m^2 \xi^2}{|b|^2}\times\nn\\
\Bigg(&
\left[
-\frac{4 \beta^5 \left(2880 y^8-62280 y^6-175507 y^4-78256 y^2-2647\right)}{15 \left(y^2-1\right)^3}
\right]_{\rm con.}\nn\\
&+
\left[
-\frac{4 \beta^5 y \left(-120 y^8+2580 y^6+8414 y^4+9325 y^2+855\right)\arccosh(y)}{\left(y^2-1\right)^{7/2}}
\right]_{\rm con.}\nn\\
&+
\left[
\frac{40 \beta^5 y \left(10 y^4-21 y^2+8\right)}{\left(y^2-1\right)^{3/2}}\right]_{\rm r.r.}\nn\\
&-\left[\frac{120 \beta^5 y^2 \left(4 y^4-8 y^2+3\right)\arccosh(y)}{\left(y^2-1\right)^2}
\right]_{\rm r.r.}
\Bigg).
\end{align}
\bibliographystyle{JHEP}
\bibliography{KinematicAlgebra}
\end{document}